\documentclass[applsci,article,accept,pdftex,moreauthors]{Definitions/mdpi} 
\firstpage{1} 
\pubvolume{15}
\issuenum{16}
\articlenumber{8916}
\pubyear{2025}
\copyrightyear{2025}
\externaleditor{Mark King} 
\datereceived{27 June 2025} 
\daterevised{24 July 2025} 
\dateaccepted{05 August 2025} 
\datepublished{13 August 2025} 
\hreflink{https://doi.org/10.3390/app15168916} 

\usepackage[final]{changes}
\usepackage{marginnote} 
\definechangesauthor[name={Second revision}, color=blue]{R2}

\Title{Forecasting the Future Development in Quality and Value of Professional Football Players}

\TitleCitation{Forecasting the Future Development in Quality and Value of Professional Football Players}

\Author{Koen 
 van Arem $^{1,}$*\orcidA{}, Floris Goes-Smit $^{2}$ and Jakob S\"ohl $^{1}$}

\AuthorNames{Koen van Arem, Floris Goes-Smit and Jakob S\"ohl}

\isAPAStyle{%
       \AuthorCitation{Van Arem, K., Goes-Smit, F., \& S\"ohl, J.}
         }{%
        \isChicagoStyle{%
        \AuthorCitation{Van Arem, Koen, Floris Goes-Smit, and Jakob S\"ohl.}
        }{
        \AuthorCitation{Van Arem, K.; Goes-Smit, F.; S\"ohl, J.}
        }
}

\address{%
$^{1}$ \quad Delft Institute of Applied Mathematics, Delft University of Technology, 2628 CD Delft, The Netherlands; \ 
\\
$^{2}$ \quad Data Science \& Computer Vision, SciSports, 3703 HX Zeist, The Netherlands}

\corres{Correspondence: k.w.vanarem@tudelft.nl}

\abstract{Transfers in professional football (soccer) are risky investments because of the large transfer fees and high risks involved. Although data-driven models can be used to improve transfer decisions, existing models focus on describing players' historical progress, leaving their future performance unknown. Moreover, recent developments have called for the use of explainable models combined with \added[comment={General Comment 3 (Reviewer 2)}]{methods for} uncertainty quantification of predictions\added[comment={General Comment 3 (Reviewer 2)}]{ to improve applicability for practitioners}. This paper assesses explainable machine learning models \replaced{in a practitioner-oriented way}{based on predictive accuracy and uncertainty quantification methods} for the prediction of the future development in quality and transfer value of professional football players. \added[comment={General Comment 3 (Reviewer 2)}]{To this end, the methods for uncertainty quantification are studied through the literature.} The predictive accuracy is studied by training the models to predict the quality and value of players one year ahead\added[comment={Comments on lines 14-23 \& lines 58-59 (Reviewer 2)}]{, equivalent to one season}. This is carried out by training them on two data sets containing data-driven indicators describing the player quality and player value in historical settings. \replaced{In this paper}{In general}, the random forest model is found to be the most suitable model because it provides accurate predictions as well as an uncertainty quantification method that naturally arises from the bagging procedure of the random forest model. Additionally, this research shows that the development of player performance contains nonlinear patterns and interactions between variables, and that time series information can provide useful information for the modeling of player performance metrics. The resulting models can help football clubs make more informed, data-driven transfer decisions by forecasting player quality and transfer value.}


\featuredapplication{This paper studies what models are most suitable for forecasting future values of player performance metrics in association football (soccer). The resulting forecast statistics find applications in team management and player scouting at football clubs. As transfer decisions concern whether a player should play for a club in the future, the predictions of future performance metrics offer a forward-looking improvement over the traditional backward-looking assessments.}

\usepackage{textcomp}

\begin{document}


\section{Introduction}\label{section: introduction}
Transfers in professional football (soccer) are a risky business because the average transfer fee has increased in recent years \citep{YangKoenigstorferPawlowski2024} and because these fees can be characterized as investments with high risks where large fees are involved \citep{McHaleHolmes2022}. Extensive knowledge about players is beneficial in making well-informed decisions about these complex transfer investments in football. \replaced[comment={Comment 1 (Reviewer 1)}]{By offering a practitioner-oriented study on forecasting players' future quality and monetary value, this paper offers methods to football clubs for gaining new insights and for improving their strategic transfer investments.}{This paper shows how football players' development in quality and monetary value can be predicted to enhance decision-making.}


Models providing information about player quality and value have recently emerged with the evolution of data-driven player performance indicators. Improvements in data-capturing technologies resulted in large data sets containing in-game data about football players, which provide the opportunity to obtain more complex variables on player performance \citep{Rein2016, HeroldGoes2019}. Numerous player performance indicators have been introduced since then. An example is the expected goals (xG) indicator, which values shot chances and shooting \mbox{ability \citep{Green2012, EggelsvanElkPechenizkiy2016, AnzerBauer2021, MeadOHareMcMenemy2023}.} Next to such action-specific models, assessment methods exist for general player performance, which can be divided into bottom-up and top-down \mbox{ratings \citep{HvattumGelade2021}.} Expected threat (xThreat) \citep{Rudd2011, VanRoy2020, VanArem2024} and VAEP \citep{Decroos2019, Decroos2020, VanHaaren2021, MendesNeves2022} are examples of bottom-up ratings that quantify action quality and use the quality of the actions to create general ratings. Top-down ratings such as plus--minus ratings \citep{SeaboHvattum2015, KharratMcHalePenaLopez2020, PantusoHvattum2021, Hvattum2020, DeBaccoWangBlei2024}, Elo ratings adjusted for team sports \citep{WolfSchmittSchuller2020}, and \added{the} SciSkill algorithm \citep{SciSkill2020} distribute credit of player performance based on the result of a team as a whole. For the monetary value of players, many models about the estimation of transfer fees and market values have been introduced and provide indicators of the current value of football players \citep{FranceschiBrocardFollerGouguet2023}. These performance and financial models describe the quality and monetary value of a football player, and \replaced{they}{these} can complement traditional scouting reports. This allows managers and technical directors of football clubs to make better-informed transfer decisions.

These models for player quality and transfer value give information about the quality and financial value of football players up to that moment, although a transfer decision regards whether a football player should be part of a team in the future. To make better-informed transfer decisions, team managers and technical directors also need insights into the \added{future} development of the indicator values that describe the financial value and \added{the }player performance\deleted{ in the future}. This paper examines the training of supervised learning models that forecast the development in player quality and transfer value of players\deleted[ comment={Comments on lines 14-23 \& lines 58-59 (Reviewer 2)}] {one year ahead}.



To this end, two prediction problems are studied: forecasting the quality of a football player and \deleted{forecasting }their transfer value one year ahead. In the first prediction problem, models are trained to predict the development of a top-down quality indicator, the SciSkill \citep{SciSkill2020}. The second prediction problem concerns the prediction of the development of the player value, described by the Estimated Transfer Value (ETV) \citep{ETV2020}. A forecasting horizon of one year is selected in this study as it aligns with the length of one full season in football\added[comment={Comments on lines 14-23 \& lines 58-59 (Reviewer 2)}]{ and the success of a transfer is often measured by the performances in the subsequent season}. The resulting models of these prediction problems offer insight into the question of whether a player will be worth the money in the future. These models thus provide critical insights for the managerial staff of a professional football club.

To further improve the usability of the research results for staff at football organizations, the findings of models should be presented such that they can be utilized by football clubs in practice, as stressed by \citet{HeroldGoes2019}. This means that models should not only be assessed on predictive accuracy but also on explainability and \added{on} methods for uncertainty quantification \citep{DaviesBransenDevos2024}. To this end, only explainable supervised learning models are used in this research, and the models are assessed on their methods for uncertainty quantification. \added[comment={Comment 2 (Reviewer 1) \& General Comment 4 (Reviewer 2)}]{Although deep learning models are common in forecasting tasks, their lack of explainability makes them less suitable for application at football clubs, and they are thus excluded in this study.} To focus even more on the applicability for practitioners, the accuracy of the models is partly determined by estimating the loss values \replaced{on}{of} different groups of football players, because certain types of football players are more important from a practitioner's perspective. 

\begin{figure}[H]
    
    \includegraphics[width=0.82\linewidth]{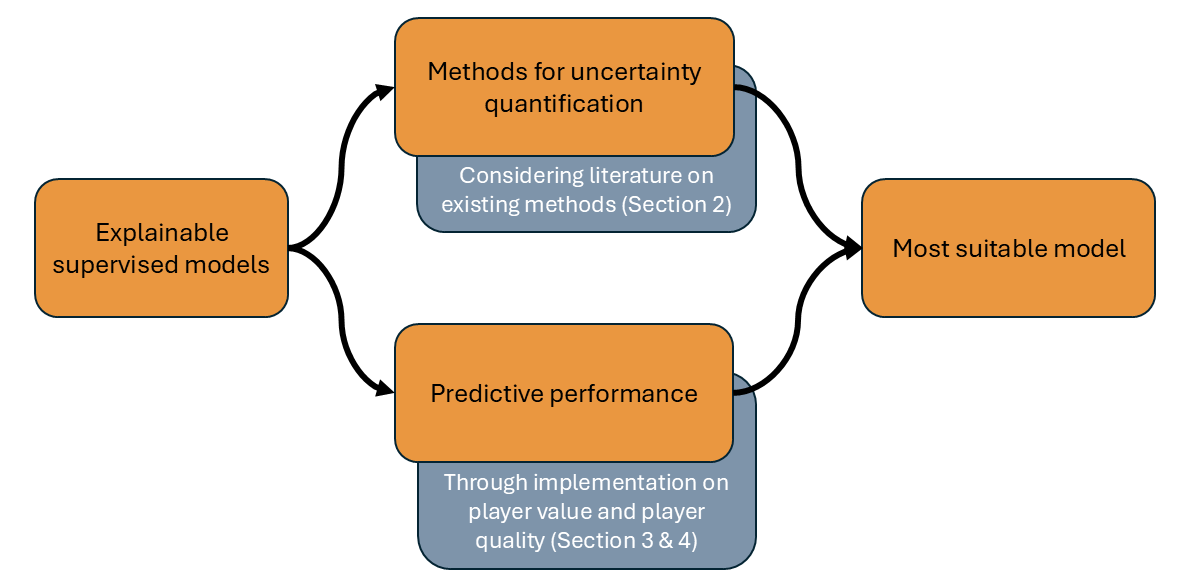}
    \caption{A visualization of the structure of this paper.\added[comment={General Comment 3 (Reviewer 2)}]{}%
    }
    \label{fig:paper structure}
\end{figure}

The aim of this research is to find the most suitable explainable machine learning model to forecast player performance with respect to predictive accuracy and methods for uncertainty quantification. To find this, several explainable supervised learning models are studied. \replaced[comment={General Comment 3 (Reviewer 2)}]{By reviewing the literature, favorable models are identified based on their methods for uncertainty quantification.}{ Their methods for uncertainty quantification are studied based on the literature.} The predictive performance is assessed by implementing them to forecast the player quality (SciSkill) and the player value (Estimated Transfer Value) of football players one year ahead. The predictive quality is studied on both the general population of players and on subgroups of players that are more interesting for practitioners, such as young or high-value players. An overview of this process is visualized in Figure~\ref{fig:paper structure}. These results will then be combined to find an answer to the research question.

The main contribution of this paper is the illustration of how extra value can be added to football player KPIs by forecasting future values \added[comment={Comment 1 (Reviewer 1)}]{in a practitioner-oriented way}. The results of this study provide knowledge on what models are most suitable for application by practitioners, lowering the threshold for real-life implementation. By taking explainability, uncertainty quantification, and performance on important subpopulations of players into account, this research is presented such that the results can be used easily by practitioners. In this way, this paper also contributes to bridging the gap between academic research and practitioners.

This paper is organized as follows: The scientific background \replaced{in the}{on} existing literature is \replaced{reviewed}{given} in Section~\ref{section: background}. \added[comment={General Comment 3 (Reviewer 2)}]{After a discussion of the considered supervised models and their methods for uncertainty quantification, favorable models are identified based on their methods for uncertainty quantification. Then, the literature on data-driven player value and on quality quantification methods is covered, along with research on forecasting these values.} The long-term forecasting of player development is then studied for the two prediction problems, for which the methods are described in Section~\ref{section: methods}. The results of the models in the prediction problems of the player quality and player value are presented in Section~\ref{section: results}. The conclusions of this research are then summarized in Section~\ref{section: conclusion}, followed by a discussion of the implications and future directions in Section~\ref{section: implications and future directions}.

\section{Background}\label{section: background}
First, the considered models are discussed along with an assessment of their methods for uncertainty quantification. Then, more background is given on existing models for football player quality and value. 

\subsection{Supervised Models}
Supervised learning models provide a possibility to forecast the player performance indicators in the future. Based on input variables, they predict output variables, and they can be trained based on an existing data set for prediction problems. Such a data set $\mathcal{D_\text{train}} = \{(X_1, y_1), \ldots, (X_n, y_n)\}$ consists of features $X_i$ describing characteristics about a data point and labels $y_i$ describing the actual values. In this research, the features $X_i$ are values that describe the situation of a player at a specific point in time, and the labels are the player's performance indicator exactly one year later. By constructing these features and labels for multiple points through time and for different players, a data set can be created to train the models.

\subsubsection{Considered Models}
To align the research with practitioner needs, explainable models are studied in this paper. The considered models can be divided into linear models, tree-based models, and kNN-based models, and they will now be shortly described.

The first linear model considered is ordinary least squares (OLS), sometimes called (multiple) linear regression, as described by (Section~3.2, \citet{Hastie2009}). This model assumes that the true relation is linear $y_i=X_i\beta+\varepsilon_i$, where $\varepsilon_i$ is normally distributed noise. OLS minimizes the residual sum of squares 
\begin{linenomath}
\begin{equation*}
RSS(\beta) = (y-X\beta)^T(y-X\beta),
\end{equation*}
\end{linenomath}
and has the lowest mean square error of all unbiased estimators. Lasso regression \cite{Tibshirani1996} is another linear model that introduces a bias in the model by minimizing the penalized residual sum of squares $RRS(\beta)+\lambda \sum_{i=1}^p|\beta_i|$. By introducing this bias, it reduces the variance, and it is possible to provide more accurate predictions. The third linear model is the linear mixed effects (LME) model, which assumes that 
\begin{linenomath}
\begin{equation*}
y_i|b_i = X_i\beta+Z_ib_i+\varepsilon_i,
\end{equation*}
\end{linenomath}
where $\varepsilon_i\sim N(0,\sigma^2\Lambda_i)$ \cite{LindstromBates1988}. The random variable $Z_i$ is called the random effect of group $i$ and can be used to model the influence of a specific random attribute that should not influence the predictions, e.g., nationality. Because of their linearity, these models are explainable and suitable for applications.

The CART decision tree (Section 9.2, \cite{Hastie2009}) is a model that recursively splits the feature space by taking splits of the form $\{X|X_j\leq s\}, \{X|X_j>s\}$ such that it minimizes the sum of squares within each split. As described by \citet{Hastie2009} (p. 312), decision trees generally suffer from a high variance. A random forest model solves this by repeatedly training decision trees on resampled data (Chapter~15, \cite{Hastie2009}). The random forest model then predicts by taking the average prediction of these decision trees. Boosting is another way of improving decision trees. A boosting algorithm fits decision trees sequentially and reweighs the data points for which the model has a bad performance (\mbox{Chapter~10, \cite{Hastie2009}}). The XGBoost \cite{Chen2016} combines several techniques, such as regularization and feature quantile estimates, to provide such a boosting algorithm. Generally, the splitting behavior in tree-based algorithms mimics the if--then reasoning of humans. Additionally, they can provide feature importances by considering how much a specific feature improves the predictions on the training set. This makes the tree-based models explainable and thus suitable models.

The last type of model considered in this paper is the k-nearest neighbor (kNN) model. Given the features, these models search for the most similar data points in the training data set and predict the associated $y_i$-value by taking the average of the $k$-nearest neighbors (Subsection~2.3.2, \cite{Hastie2009}). The closest neighbors can be weighted more heavily as illustrated by \citet{Dudani1976}. The ReliefF algorithm by \cite{RobnikSikonjaKononenko1997} calculates feature importances by looking at the probability of having different values considering the neighbors of a data point. \deleted{A downside of the kNN models is that they suffer from the curse of dimensionality as illustrated in \cite[Section~2.5]{Hastie2009}. Nonetheless, the kNN models can provide examples of training data points to explain predictions. In the context of the current study, this would be players who have been in similar situations in history, which gives a very intuitive interpretation to practitioners.}
\added[comment={Comment 3 (Reviewer 1) \& General Comment 2 (Reviewer 2)}]{Additionally, the kNN models can provide examples of training data points to explain predictions. This gives an intuitive interpretation to practitioners, since they can see that a player’s forecast performance is based on previous players in similar situations, making the rationale behind the prediction clear and relatable. To interpret a prediction, practitioners can then investigate the players in similar situations, which can help them assess the individual prediction.} \added[comment={Comment on lines 116-156 (Reviewer 2)}]{However, similarity becomes an abstract concept for practitioners in higher dimensions (pp.~108--109, \cite{Molnar2019}), and the kNN model suffers from the curse of dimensionality (Section~2.5, \cite{Hastie2009}), which makes application with a large feature set infeasible. To mitigate this in the current paper, kNN models are applied to a feature set of limited dimensions.}

\subsubsection{Uncertainty Quantification}
To better align research with application in practice, the models in this study are also assessed on their methods for uncertainty quantification. Although quantile regression could be applied to obtain a form of uncertainty quantification, this requires an extra model. This makes quantile regression less useful for applications in practice. Therefore, models with uncertainty quantification methods that do not need to alter the model or results are favorable.

Linear regression has an underlying theory that gives prediction intervals as described by \citet{NeterKutnerNachtsheimLi2004}. These prediction intervals are based on assumptions that are unlikely to hold for the prediction problems in this paper, such as the normality of errors. Nonetheless, they do give an indication of the uncertainty of the prediction. Similarly, the kNN models provide the neighbors, which are a group of similar data points. Uncertainty quantification can be obtained by taking the minimal and maximal values of the dependent variable within these neighbors if the number of neighbors $k$ is of significant size. Lastly, the bagging procedure of the random forest model can be utilized to obtain uncertainty quantification for the predictions, as described by \citet{WagerHastieEfron2014}. This method utilizes the different decision trees in the random forest to quantify the uncertainty in the prediction. From these properties, it is concluded that the linear regression, random forest, and kNN-based models are favorable with respect to uncertainty quantification.

\subsection{Existing Indicators}
In recent years, the increasing amount of available data in football has driven the introduction of methods to rate individual football players \citep{ArntzenHvattum2021}. The performance of a professional football player has traditionally been determined via expert judgment based on video data and statistics describing the frequency of in-game events.
Data-driven models have offered the possibility to reduce the bias in assessments and improve the consistency of the judgment of both player quality and player value, as discussed below. In this way, these models have provided new and consistent insights into the quality and monetary value of football players, which has aided in transfer decisions.

\subsubsection{Player Performance}
The creation and the comparison of models for player performance have revealed new challenges. Because teams can have different aims in football and can apply various tactics, there does not exist a ground truth for player performance \citep{DaviesBransenDevos2024}. A player can, for instance, be instructed to keep the ball in possession, which leads to the player performing fewer actions that might result in scoring a goal. Because of this lack of ground truth, and various models exist that describe different aspects of the game.

\paragraph{\added[comment={Comment 4 (Reviewer 1)}]{Bottom-Up Ratings}}

 Some 
 models define the players' quality by their actions\added[comment={Comment 4 (Reviewer 1)}]{, called bottom-up ratings \cite{HvattumGelade2021}}. Although action-specific models exist \citep{Green2012, EggelsvanElkPechenizkiy2016, GoesKempeMeerhoffLemmink2018,  AnzerBauer2021, MeadOHareMcMenemy2023}, methods to assess the quality of all types of actions have been introduced that generalize the action-specific models. These general models often define a `good' action as one that increases the probability of scoring and decreases the probability of conceding a goal. 

The VAEP model by \Citet{Decroos2019} calculates the probability of scoring given the last three actions, including in-game context. Because the involved machine learning techniques are considered a black-box model by practitioners, the authors in \cite{Decroos2020, VanHaaren2021} introduced methods to make the VAEP model more accessible for practitioners. 
A comparable, more interpretable framework is the Expected Threat (xThreat) model by \Citet{Rudd2011}, which only considers the current situation, defined by the location of the ball-possessing player, to estimate the probabilities of the ball transitioning to somewhere else on the pitch. This is modeled by a Markov chain, which can be used to describe the probabilities of scoring before and after each action to find the quality of an action. \Citet{VanRoy2020} have compared the xThreat and VAEP models and have found that, although the xThreat model is more interpretable to practitioners, it can only take into account the position of an action and excludes contextual information such as the position of defenders from its model. This means that there is a trade-off between explainability and the inclusion of in-game context when choosing either VAEP or xThreat models. \Citet{VanArem2024} have extended the xThreat model by including variables describing the defensive situation and height of the ball. This Extended xThreat model can take into account in-game context like the VAEP model, while maintaining explainability as an xThreat model. 

\added[comment={Comment on lines 209-212 (Reviewer 2) \& Comment 4 (Reviewer B)}]{These bottom-up ratings describe the quality of a professional football player by assessing in-game on-the-ball actions. These on-the-ball actions are often offensive actions, and they are better at capturing the quality of attackers and attacking midfielders, which leads to a bias when these models are studied. This bias makes it harder to assess the quality of defensive players. Moreover, these models need more detailed event data describing in-game events, limiting the scale on which they can be applied. Consequently, the quality of football players in this study is not assessed with bottom-up ratings.}

\paragraph{\added[comment={Comment 4 (Reviewer 1)}]{Top-Down Ratings}}
\deleted[comment={Comment 4 (Reviewer 1)}]{The bottom-up ratings like xThreat and VAEP describe the quality of a professional football player by assessing in-game on-the-ball actions. On-the-ball actions are often offensive actions, and they are better at capturing the quality of attackers and attacking midfielders, which indicates that a bias occurs when these models are studied. This bias makes it harder to assess the quality of defensive players, and it can be reduced by using top-down models that describe player quality using the lineups and outcomes.}
\added{These problems of bottom-up ratings \cite{HvattumGelade2021} can be reduced by using top-down models that describe player quality using the lineups and outcomes.} Plus--minus ratings are an example of such ratings and were first used in ice hockey and basketball \citep{KharratMcHalePenaLopez2020}, and were later applied to football by \Citet{SeaboHvattum2015}. For plus--minus ratings, the game is partitioned into game segments that contain the same lineups, which correspond to the data points in the data set. In this data set, the result of the game segment, the goal difference, for example, is the dependent variable. Indicators describing whether a player was active in the segment are the independent variables. Linear regression is then applied to estimate the influence of players on the results. The coefficients of the regression describe the average impact of a player on the game result and give an indicator for player performance over the period of time covered by the data. 

Because substitutions are infrequent in football, a game does not have many game segments with different lineups. Moreover, football is a low-scoring sport. This creates a situation where a low number of segments with limited distinction in outcomes must be used to infer player quality. To deal with this, other quantities have been used as the dependent variable to describe the result of a segment, like the expected number of goals (xG), the expected number of points (xP), and the created VAEP values by \citet{KharratMcHalePenaLopez2020} and \citet{HvattumGelade2021}. \citet{PantusoHvattum2021} additionally showed the potential of taking age, cards, and home advantage into account, and \citet{Hvattum2020} illustrated how separate defensive and offensive ratings can be obtained. Because putting a player in the lineup is an action that a coach performs, \citet{DeBaccoWangBlei2024} adapted the plus--minus ratings using a causal model to better describe the influence of the selection of a player. These studies show how plus--minus ratings have been adapted to the application of rating players in football.

Whereas the plus--minus ratings describe the quality of a player using multiple historical games, there also exist models that determine the quality of a player after each game using the lineups and final score. Elo ratings are such ratings and were originally developed to evaluate performance in one-on-one sports. The concept was subsequently adapted to the game of football by \citet{WolfSchmittSchuller2020}. This adapted algorithm provides ratings for each individual football player and calculates the team rating via the average of players in a game weighted by the number of minutes played. The ratings are then used to predict the match outcome using a fixed logistic function, and after each game, the individual ratings are adjusted. If the outcome is better than predicted, the player rating is increased, and if the outcome is worse than expected, it is decreased. The authors also introduced an indicator for player impact to deal with the fact that this rating undervalues good players at below-average teams. 

The SciSkill \citep{SciSkill2020} is \replaced[comment={Comment 5 (reviewer 1)}]{an industry-validated, more elaborate version of the Elo rating}{a more detailed Elo rating}. Instead of only considering one player's quality, it describes football players combining a defensive rating and an offensive rating. These offensive and defensive scores are then combined to obtain one value, the SciSkill. For each game, the outcome \deleted[]{of the game} is predicted using a model via an expectation-maximization algorithm. After the prediction, the SciSkill is updated by adjusting the SciSkill values based on the difference with the actual game result \added[comment={Comment 5 (reviewer 1)}]{as with other Elo-ratings}.
Compared to the Elo algorithm\added[comment={Comment 5 (reviewer 1)}]{, which was implemented and validated by \citet{WolfSchmittSchuller2020}}, the SciSkill model extracts more detailed information from the matches and describes the player quality more elaborately. 

\subsubsection{Player Value}
In contrast with player quality, there does exist a ground truth for the concept of player values. The value of a transfer fee is based on the value of the player for each of the involved clubs and historical transfers of similar players \citep{PoliBessonRavenel2022}. Thus, the historical values of these fees can be used to predict the transfer value of a player, albeit at the cost of some selection bias.

A well-known medium that describes the value of a player is Transfermarkt. This company uses crowd estimation to assign values to players \citep{HermCallsenBrackerKreis2014}. These market values describe the general value of a player and do not take into account the temporary situation of a football player, like the current club and contract length. This means that they describe a different quantity than the expected value of a transfer fee. Nonetheless, these values are strongly correlated with the real transfer fees as shown by \citet{HermCallsenBrackerKreis2014}. Therefore, both market values by Transfermarkt and transfer values are often used interchangeably when describing the monetary value of a football player. 

The literature review by \citet{FranceschiBrocardFollerGouguet2023} showed that many studies have been performed to describe the monetary value of a football player based on data. The authors found that most of these studies used linear regression to find which variables have a significant linear dependence on the value of a football player. The review by \citet{FranceschiBrocardFollerGouguet2023} considered 111 trained models that were used in the scientific literature to investigate the transfer value of a football player. The vast majority (85\%) of the models are based on ordinary least squares. The authors have also shown the importance of different variables in the considered models. For instance, they have found that age, the square of the age, and the number of matches played by a player are frequently studied variables. These variables are also most often found to be significant. Players frequently increase in value as they get better with experience that is gained over the years, but they also decrease in value as a player loses the potential to improve when getting older. Consequently, the influence of age is often measured with a quadratic term. Similarly, the number of games played can be expected to be an important variable because players gain experience by playing games, which makes them more valuable. Moreover, players who play a lot of games are often the better players on a team. This explains why these variables are important, as found by \citet{FranceschiBrocardFollerGouguet2023}. Their study also shows that almost no variables describing defensive behavior were considered when other researchers trained models to describe transfer values. As a consequence, the resulting models can be expected to describe the value of offensive players better than that of defensive players. In addition, the linear models in these studies are mostly trained to determine the influence of variables on the transfer value of a football player, and they are generally not trained and tested for out-of-sample prediction, which limits the application of predicting based on new, unseen data. 

In contrast, \citet{AlAsadiTasdemir2022} have trained multiple models to predict market values of football players with features from the video game FIFA. They have found that a random forest model gives improved predictions over linear methods. \citet{ArrulSubramanianMafas2022} have studied the application of artificial neural networks for the same problem, also considering features from the video game FIFA. They obtain similar loss values as the random forest model of \citet{AlAsadiTasdemir2022}. The research by \mbox{\citet{BehravanRazavi2021}} introduces a methodology to train a support vector regression model via particle swarm optimization for the prediction of market values. Although the data set is somewhat similar to the studies above, this model attains worse loss values. In the study of \mbox{\citet{YangKoenigstorferPawlowski2024}}, random forests, GAMs, and QAMs are applied to predict the transfer fees of players based on variables describing the player. The authors have inferred from their random forest model that the expenditure of the buying club and the income of the selling club are important features in predicting the transfer fee, as well as the age and the remaining contract duration. This research additionally illustrates how GAMs and QAMs can be used to investigate the dependency between the player transfer value and the given features. The QAM models show that this relation varies for different quantiles, indicating the need to study the influence of models on different groups of players. On the other hand, the GAMs show that the relationship between the transfer values and the features is often nonlinear. A study with extensive types of player performance metrics as features has been performed by \citet{McHaleHolmes2022}, in which linear regression, linear mixed effects, and XGBoost models have been trained. These models not only include statistics such as the number of minutes played, height, and position, but also plus--minus ratings based on xG, expert ratings from the video game FIFA, and GIM ratings, which are similar to VAEP ratings. The results show that the best predictive performance is attained by the XGBoost model, although the linear mixed effects with the buying and selling clubs as random effects also provide good results. Their results indicate that their model outperforms Transfermarkt market values when predicting the transfer fees on average, although the market values are a better predictor for transfers of more than EUR 20 million. These studies show how supervised learning can be used to obtain models that predict the value of a football player. They show that nonlinear methods generally predict more accurately and that the patterns can differ for different groups of players. 

\subsection{Predicting Future Values}
Although the studies discussed above are concerned with the quantification of the quality and monetary value of football players at the present moment, only limited work has been conducted on the future development of player performance. \mbox{\citet{ApostolouTjortjis2019}} have conducted a small-scale study with the aim of predicting the number of future goals of the two football players Lionel Messi and Luis Suárez using a random forest, logistic regression, a multi-layer perceptron classifier, and a linear support vector classifier. \citet{ChazanPantzalisTjortjis2020} have predicted the expert ratings in the next season of 59 center-backs in the English Premier League based on one season of player attributes from a popular football manager simulation game. Their method uses a linear regression model to describe the in-sample patterns. \citet{GiannakoulasPapageorgiouTjortjis2023} have trained linear regression, random forest, and multi-layer perceptron models to predict the number of goals of a football player in a season before the start of the corresponding season. Their data set entails around 800 football players. Similarly, the models by \citet{MarkopoulouPapageorgiouTjortjis2024} have been trained to predict the number of goals by looking at the creation of different models per competition. This study on 424 football players shows that the best results are often obtained using XGBoost models for this prediction problem and that making different models for different competitions might be beneficial. 

\citet{BarronBallRobinsSunderland2018} have tried to predict the tier within the English first three leagues in which a football player would play next season as an indicator of player quality. Three artificial neural networks are trained to predict in which league a player would play with the data of 966 football players. Their models are only able to recognize the differences between players in the lowest and highest tiers (League One and the English Premier League). 

Little literature exists about the prediction of the future transfer values of football players. \citet{Baouan2022} have applied lasso regression and a random forest model to identify important features for the development of around 22,000 football players. They have trained these supervised models for players of different positions to predict a player's transfer value two years in the future based on performance statistics. The feature importances of the models show, for instance, that the average market value of a league is an important feature for the future values of players in that league. Although cross-validation is performed for the hyperparameter tuning, this study focuses on finding in-sample patterns.

Our research treats forecasting of both player quality and transfer value. The current paper builds on the existing literature about forecast player performance by studying the long-term forecasting of a model-based player quality indicator on a larger data set. Additionally, this research avoids the bias in the current literature that better describes offensive players by using a more general top-down rating. Our research contributes to the existing knowledge of forecasting the monetary value because the models are trained to perform out-of-sample prediction, which makes it possible to apply them to unseen situations. Moreover, the combination of forecasting the development of both player quality and transfer value gives a comprehensive summary of the most important \replaced{aspects}{factors} in transfer decisions. In this way, this paper \replaced[comment={Comment on line 370 (Reviewer 2)}]{builds upon the existing}{fills a gap in the} literature by forecasting the development of model-based indicators for player quality and value on a significant data set in a predictive setting.

\section{Methods}\label{section: methods}
The goal of this study is to find the most suitable machine learning model to forecast the development in player performance with respect to predictive accuracy and uncertainty quantification methods. The assessment of the uncertainty quantification methods was performed using the literature on the models in Section~\ref{section: background}. To compare the predictive performance, the models are trained to predict player performance indicators one year ahead in two prediction problems, and the corresponding loss values are determined. The two prediction problems concern the prediction of the development in player quality based on the top-down SciSkill rating and the development in monetary value described by the Estimated Transfer Value (ETV) model.

\subsection{Data}
\begin{figure}[H]
    
    \includegraphics[width=0.75\linewidth]{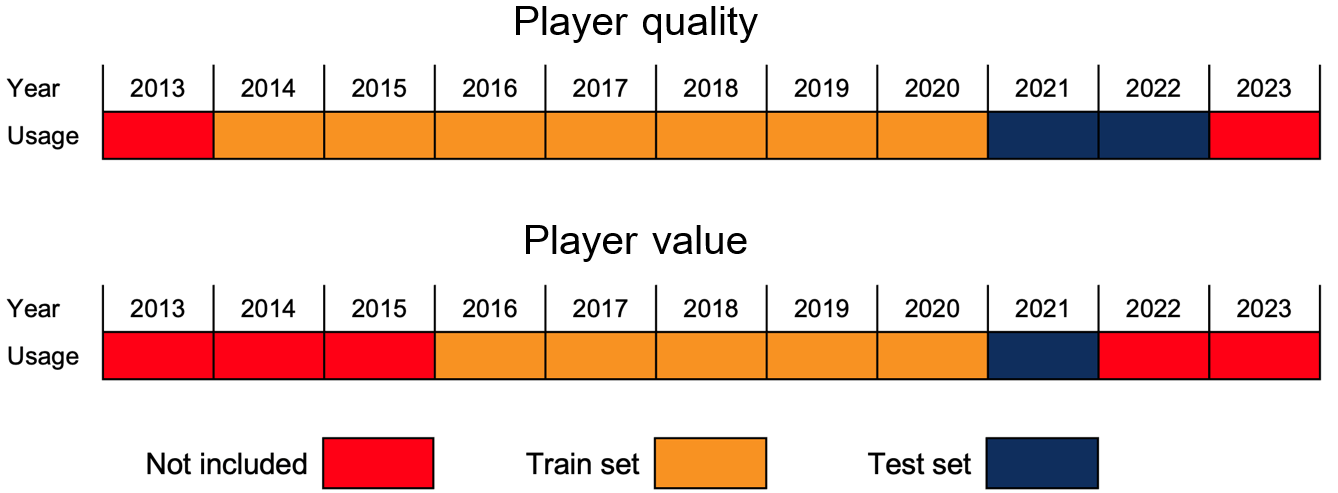}
    \caption{A visualization of the partition of the years in the test and training set in both prediction problems.}
    \label{fig: distribution years}
\end{figure}

Two data sets were obtained to study the predictive performance of the models in this paper. The features of the data points in both prediction problems represent the historical situation of professional football players. The corresponding label is the player performance indicator that was recorded one year later.
The data availability for different years is visualized in Figure~\ref{fig: distribution years}. For the player quality prediction problem, data is available from 2014 up to 2022, while the data for the player value covers the years 2016 up to 2021. 

\added[comment={Comment 6 (Reviewer 1)}]{Since the used SciSkill and ETV models are proprietary, exact reproduction of the results using the same models is not possible. However, similar models exist for both the SciSkill \cite{WolfSchmittSchuller2020} and ETV models \cite{AlAsadiTasdemir2022, ArrulSubramanianMafas2022, BehravanRazavi2021, YangKoenigstorferPawlowski2024, McHaleHolmes2022}, as discussed in Section~\ref{section: background}, ensuring that the overall approach remains, in principle, reproducible.}

\subsubsection{Player Quality: SciSkill}
The first prediction problem concerns predicting the development in the subsequent year of the player quality described by an EM algorithm called the SciSkill \cite{SciSkill2020}, which is a generalization of the Elo rating, as discussed in Section~\ref{section: background}. The data set is restructured to contain monthly data points describing the situation of each player at that time. Each monthly data point consists of the features and the dependent variable. The dependent variable is the difference between the player quality one year ahead and the current player quality value, which is the development in player quality in one year. The features set consists of 86 features constructed with domain knowledge describing, for instance, the month of year, the current player performance, league strength, time since the most recent game, player characteristics, time series information of the SciSkill, the club's transfer situation, and the difference in quality between the player and his teammates. The dependent variable and features are described in Tables~S1--S3 in the Supplementary Materials.

Only male players with more than 20 games and more than 2 years of data are considered. \added[comment={Comment on lines 387-431 (Reviewer 2)}]{This filtering might introduce a bias in the data against player groups on which less data is available, such as younger or injury-prone players. However, it is necessary to obtain enough data points per player. The results on players with limited data availability should thus be interpreted with care.} The final data set consists of 80,568 male professional football players playing in the years 2012 to 2023. As the data set consists of 3,834,539 data points, there are on average 47.6 monthly data points per player, which corresponds to roughly 4 seasons of data.

\subsubsection{Player Value: Estimated Transfer Value}
In the second prediction problem, the monetary player values are considered, obtained by the Estimated Transfer Value model (ETV). This model is a supervised tree-boosting model, trained on historical transfers to predict the transfer fees based on features that describe the situation of the player at the time of transfer. The model is then used to describe the transfer value for professional football players to obtain the transfer values for the general population of players over time. In this way, the supervised ETV model provides the monetary values of players over time.

The data set for this prediction problem of the development in player value describes the current situation of professional football players with 58 features. These features were constructed based on domain knowledge and data availability. Because there were fewer data points than for the player quality case study, the number of features was reduced by eliminating highly correlated features. These features consist of indicators of league strength, age, experience, contract situation, the current quality (SciSkill), or the monetary value (ETV) of a football player. The feature set also includes player characteristics, like playing position and age, as well as the differences in quality with teammates, the transfer history of his current club, and league strength. A description of all features and the dependent variable in this prediction problem can be found in Tables~S4--S6 in the Supplementary Materials.

For this prediction problem, the data set consists of biannual data points describing the players' transfer values in January and July within the period from 2014 to 2021. Similarly to the SciSkill, players with less than 20 games, less than 2 years of data, or missing values are excluded. 
The remaining data set includes 60,175 male professional football players described with 413,177 biannual data points. This corresponds to an average of 6.87 data points per player, equivalent to approximately 3.5 seasons of data.  

\subsection{Model Implementation}
\subsubsection{Model Assessment}
Supervised models are trained on the prediction problems of player quality and player value. The root mean square error (RMSE) and mean absolute error (MAE) are determined for the test set. The RMSE is more strongly influenced by large errors, which can be expected to happen for `superstar' players. Since these players are especially of interest to practitioners, the performance in the RMSE is considered the most important. 

The losses are determined on different parts of the test set. First, the loss values of the RMSE and MAE are calculated using the complete test sets. Second, the loss function of the RMSE is also considered for different age groups on the test sets because estimating the potential of a player is mostly interesting for young players. Third, the RMSE is studied on important subgroups of players, like players with large positive or negative development, players with good performance, or players with a high transfer value. By determining the test losses separately on the general population of players, young players, and important subgroups of players, the models are studied on their predictive performance.

Beforehand, 5\% of all football players are left out of both data sets based on stratified sampling for internal studies by the data provider. Because the data of player performance indicators is often dependent on time \citep{DaviesBransenDevos2024, MendesNeves2022}, time-dependent train--test splits are applied to study the predictive performance on unseen data, as visualized in Figure~\ref{fig: distribution years}. In both case studies, all data up to 2020 is considered as the training set, and from 2021 and later as the test set.

\subsubsection{Model Training}
In this study, linear, tree-based, and kNN-based models are implemented. The linear models are ordinary least squares (OLS), also known as multiple linear regression (MLR), lasso regression, and a linear mixed effect model (LME). For the OLS model \citep{Seabold2010}, feature selection is performed by applying backwards selection with a threshold of the p-values of 0.0001 for the player quality and 0.001 for the player value. These values \deleted[comment={Comment 7 (Reviewer 1)}]{have been determined via trial and error, considering values of $10^k$ for multiple values of $k$, and they} are smaller than commonly used for significance testing\deleted{. This is} due to the predictive nature of this study and the large data sets combined with the fact that common assumptions of the linear regression model, like normality, do not hold. \added[comment={Comment 7 (Reviewer 1)}]{These p-values were selected via trial and error, considering values of the form $10^k$, such that roughly half of the features is excluded.} For lasso regression \citep{Pedregosa2011}, feature selection is applied by selecting only the variables with nonzero coefficients. The linear mixed effect (LME) model \citep{Seabold2010} is trained to take into account the influence of a player's nationality as a random influence. The feature selection for the LME model is performed by taking the 20 variables corresponding to the largest feature importances within the lasso model. This is conducted for computational feasibility. For these three linear models, no interaction effects are included.

The tree-based models are the decision tree \citep{Pedregosa2011}, random forest \citep{Pedregosa2011}, and \mbox{XGBoost \citep{Chen2016}} models. For the tree-based models, feature selection is performed by first adding noise variables to the data set and training the models. All features with a larger feature importance than the noise variables are then selected. As tree-based methods can have a bias favoring non-discrete variables, both discrete and continuous variables are added. The discrete features are compared with the discrete noise variables, and the continuous features with the continuous noise variables.

To investigate the predictive power of time series, three k-nearest neighbors (kNN) models have been implemented using the Hierarchical Navigable Small Worlds indexer provided by \citet{Douze2024}. This was carried out on an altered feature set consisting of time series information with lagged versions of the most important player indicators. Because the kNN model suffers from the curse of dimensionality (Section~2.5, \cite{Hastie2009}), the feature set is kept relatively small. The features and labels of this altered data set are described in Tables~S3 and S6 in the Supplementary Materials. First, a normal kNN model is applied to the predictive problem. This model searches for the most similar data points with respect to the Euclidean norm. The prediction is then obtained by applying a weighted average, where closer data points are weighted more heavily as described by \citet{Dudani1976}. Possible weighting methods are the reciprocal of the absolute value of the distance, the distance with min--max scaling, or uniform weights. The method of calculating these weights is considered a hyperparameter. The second kNN model is constructed \replaced[comment={Comment 8 (Reviewer 1)}]{in the same way as the first}{similarly}, but it calculates the distances based on the Mahalanobis distance\added[]{ instead of the Euclidean distance}. The Mahalanobis distance projects the features onto a decorrelated feature space and calculates the Euclidean distance \added[]{in this changed feature space}. This makes it possible to better distinguish differences because the lagged time series features are heavily dependent. Lastly, an adapted RReliefF method \citep{SikonjaKononenko2003} is implemented to calculate feature importances of the normal kNN model in a regression context. The features are then multiplied by the feature importances before calculating Euclidean distances to introduce new feature weights. The kNN model with Euclidean distance is then trained on these reweighted features.

\subsubsection{Hyperparameter Tuning}
\begin{figure}[H]
    
    \includegraphics[width=0.7\linewidth]{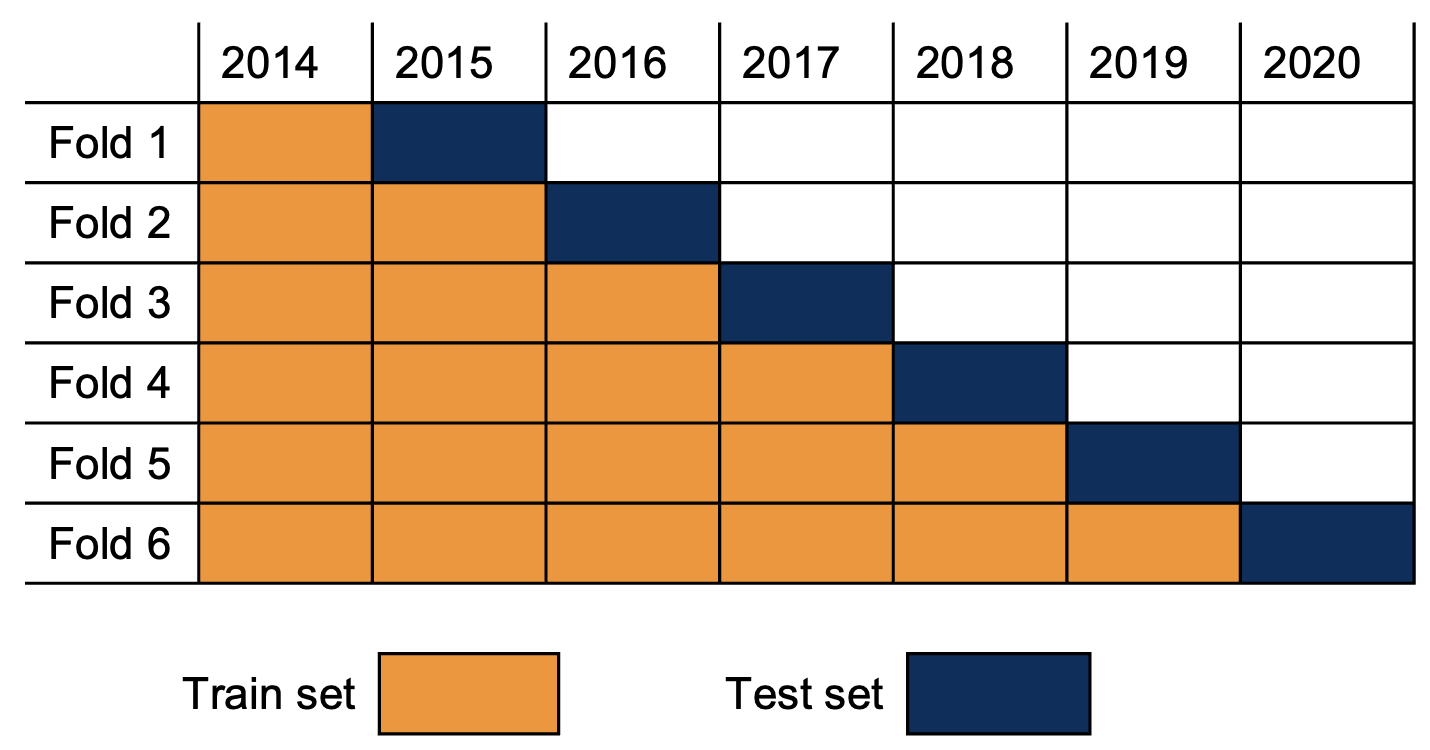}
    \caption{A visualization of the distribution of the data points in the test and training set for the player quality prediction problem of the adjusted version of cross-validation used for hyperparameter optimization.}
    \label{fig:adjusted_cross_validation}
\end{figure}

To select the best hyperparameters of the models in the case studies, Bayesian optimization \citep{Louppe2016} is applied to find hyperparameters minimizing the RMSE loss. Because of the time dependencies of player performance indicators, \replaced[comment={Comment on lines 503 (Reviewer 2)}]{cross-validation with an expanding windows split was implemented. This time series split strategy}{an adjusted version of cross-validation is implemented that} splits \added[]{the data} per year and incrementally grows the training set as visualized for the player quality prediction problem in Figure~\ref{fig:adjusted_cross_validation}. After the hyperparameter tuning, the models are trained on the complete training set with the optimized hyperparameters.

\subsection{Feature Importances}
To illustrate the advantages of using explainable models, the feature importances are studied in both predictive problems. Among the models in this paper, the linear and tree-based models have methods that can be used to calculate the feature importances of the models. To investigate what features are important for the development of professional players, the feature importances are calculated. This is similar to the methods by \citet{Baouan2022}. However, because the development of the indicator is considered instead of the indicator value itself, it is possible to describe the influence of the indicator value on the development. Min--max scaling is applied to the feature importances to be able to compare them across the different models.

\section{Results}\label{section: results}
\subsection{Prediction Problem on Player Quality}\label{res:prediction problem player quality (sciskill)}
\subsubsection{General Population of Players}
\begin{figure}[H]

{\captionsetup{justification=centering}
\begin{adjustwidth}{-\extralength}{0cm}
\centering

\subfloat[\label{fig:results:sciskill:general:rmse}]{
    \includegraphics[width=8.5cm]{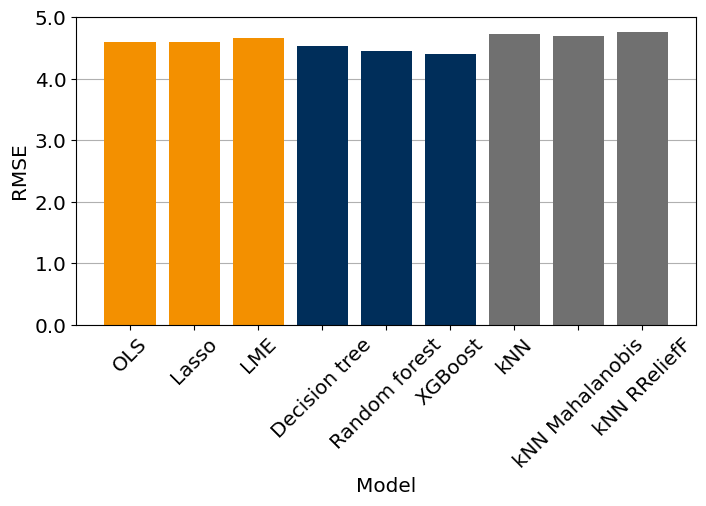}
}
\hfill
\subfloat[\label{fig:results:sciskill:general:mae}]{
    \includegraphics[width=8.5cm]{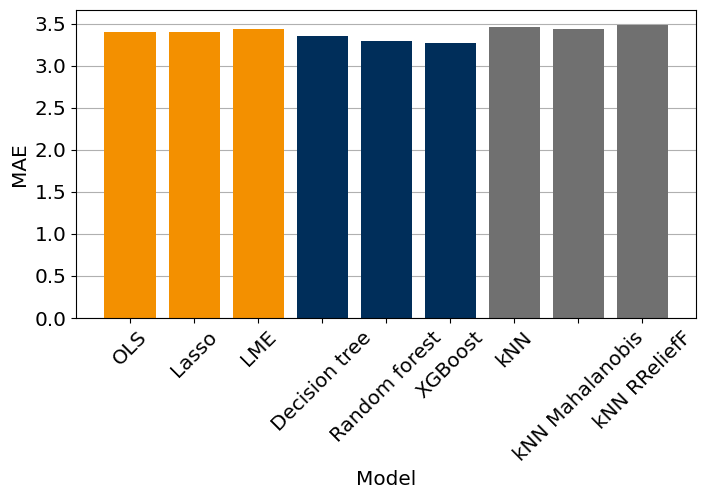}
}

\end{adjustwidth}}
\caption{
The test loss values for the different models on the general population of players in the player quality prediction task for two different loss functions: 
(\textbf{a}) RMSE and
(\textbf{b}) MAE.
}
\label{fig:results:sciskill:general results}
\end{figure}

The loss values on the test set are shown in Figure~\ref{fig:results:sciskill:general results}. The RMSE and MAE agree on the predictive performance of the models to forecast player development in general player quality, as they show similar patterns. The results indicate that the XGBoost model attains the lowest loss values \added{of all models.} \replaced[comment={General Comment 1 (Reviewer 2)}]{The decision tree and random forest models also attained low loss values, with the latter having slightly better loss values.}{ with the random forest having the second-best losses. }

The tree-based models generally obtain the lowest loss values. The difference in performance with the linear models implies that there is some nonlinear or interaction effect in the true underlying relation. On the other hand, the kNN models based on the time series attain the worst loss values. This implies that these kNN-based models missed out on important information by relying on time series information with a local method. It shows that the development of player quality is nonlinear and dependent on contextual information.

%
%
%

\subsubsection{Predictions per Age}
\begin{figure}[H]
    
    \includegraphics[width=0.99\linewidth]{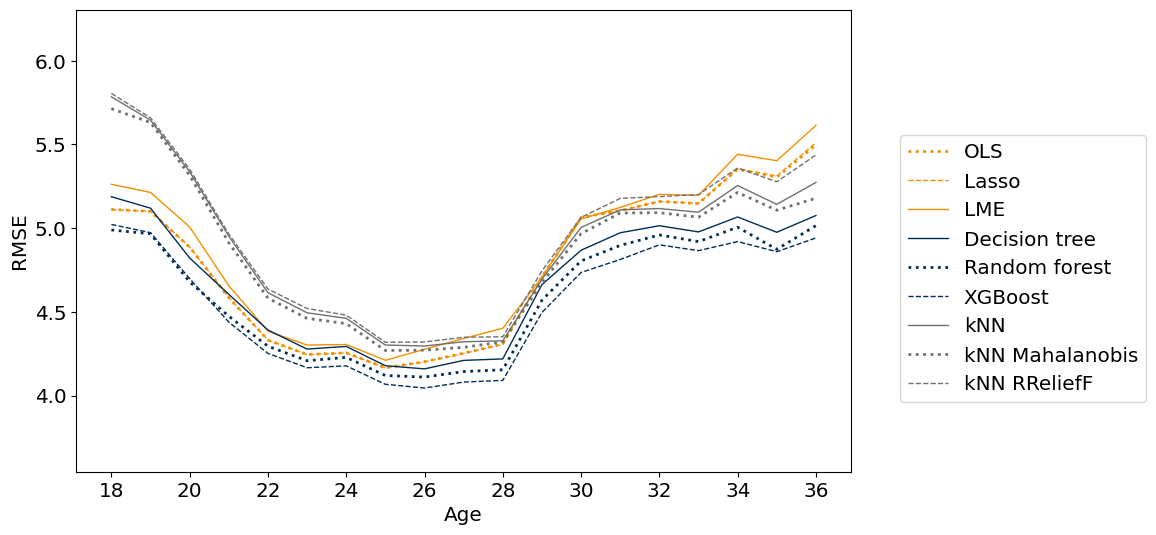}
    \caption{The RMSE values for each age for the different models in the player quality prediction problem.}
    \label{fig:results:sciskill:error per age}
\end{figure}

The RMSE loss is also determined for the players of each age in the test set, as visualized in Figure~\ref{fig:results:sciskill:error per age}. \deleted[comment={Comment 9 (Reviewer 1)}]{The XGBoost and random forest models perform better for all ages. For ages above 25, the XGBoost model has better loss values compared to the random forest model, whereas the differences are less evident for the lower ages. Because the younger ages are most important for estimating the potential of players in practice, these results indicate that both the XGBoost and random forest models are favorable models.} In general, Figure~\ref{fig:results:sciskill:error per age} shows that all models predict best on players with ages 24 up to 28. The quality of younger or older players is more volatile because young players tend to increase in quality, and older players tend to decrease in quality\added{, albeit over a longer time period}. Additionally, more data is available on players around their peak age. \added[comment={Comment 9 (Reviewer 1)}]{The data set consists of roughly 10,000 data points for players aged 18 and 20,000 for players aged 34, while around 70,000 data points describe situations of players aged 25.} The worse predictive performance on the younger and older players \replaced{can be}{is} explained by the lower number of data points combined with a higher volatility in player quality. \added[comment={Comment 9 Reviewer 1}]{Moreover, the low number of data points for young players indicates that the RMSE estimates for young players may be less stable than those of players aged 24 up to 28.}

\added[comment={Comment 9 (Reviewer 1)}]{The results show that the XGBoost model attains the lowest loss values for almost all ages, especially for ages above 22. For the ages below 22, the random forest model attains similar loss values. \deleted[comment={General Comment 1 (Reviewer 2)}]{The XGBoost and random forest models perform better for all ages. For ages above 25, the XGBoost model has better loss values compared to the random forest model, whereas the differences are less evident for the lower ages.} Because the younger ages are most important for estimating the potential of players in practice, these results indicate that both the XGBoost and random forest models are favorable models.}

Additionally, the results show that the kNN models perform worse on young players. \added[comment={Comment on lines 527-539 (Reviewer 2)}]{Generally, younger players are harder to model because of their volatility. Combined with the fewer available data points, this explains why the local kNN models can predict the development of younger players less well.} \added{Conversely,} the linear methods tend to perform worse for older players. This indicates that \deleted{nonlocal methods with more features beyond time series perform better for young ages, whereas} the patterns for older players involve nonlinearities or interaction effects. \added{Because tree-based models are non-local methods that can take into account nonlinearities and interaction effects, they do not suffer from these problems. } This explains the low loss values of the tree-based models on the general test set as shown in Figure~\ref{fig:results:sciskill:general results}a,b.


\subsubsection{Prediction on Important Player Groups}
The test losses are determined on three different groups of players that are important for the application of the models in this prediction problem: high-quality players, players with a large decrease in performance, and players with a large improvement in performance. \added[comment={Comment 10 (Reviewer 1)}]{Based on domain knowledge, high-quality players were defined as players with a SciSkill of more than 100, while an increase or decrease of more than 10 was defined as large.} The RMSE values on the test set for these subgroups are shown in Figure~\ref{fig:results:sciskill:subsets}. In general, the vertical scales show that the RMSE on these subgroups is larger than that of the general population. This is even more evident for players with a large increase, and it indicates that the development of these players of interest is harder to predict. 

%
%

%
%

The results also show that the tree-based models are best at predicting the development of all three of these subgroups of players. The \deleted[comment={General Comment 1 (Reviewer 2)}]{random forest and} XGBoost model\deleted{s appear to predict} \added{attains} the best \added{RMSE values} for \replaced{all three of the important groups of players.}{the high-quality players and the players with a large decrease in quality.} Figure~\ref{fig:results:sciskill:subsets}c shows that the XGBoost model outperforms all other models on the group of players with a large increase in quality\added[comment={General Comment 1 (Reviewer 2)}]{, whereas the differences in loss values with the decision tree and the random forest are less apparent in Figure~\ref{fig:results:sciskill:subsets}a,b}. \deleted{The random forest model attains the second-best loss values.}

The results indicate that the linear models predict significantly less accurately for the group of players with a large decrease in quality compared to the other models. This shows that interaction effects or nonlinearities are particularly important for predicting a decrease in player quality. On the other hand, the kNN models based on time series predict significantly worse for players with large increases in quality, as the large increases are probably for the young players. This is in line with Figure~\ref{fig:results:sciskill:error per age}, where kNN methods do not perform well on the young players. These differences in losses show that the large increases in player quality are better predicted by global models with contextual information and that decreases are better predicted by methods that include nonlinearities or interaction effects. 

\subsubsection{Feature Importances}
\begin{figure}[H]
    \centering
    \begin{adjustwidth}{-\extralength}{0cm}
    \includegraphics[height=0.92\textheight]{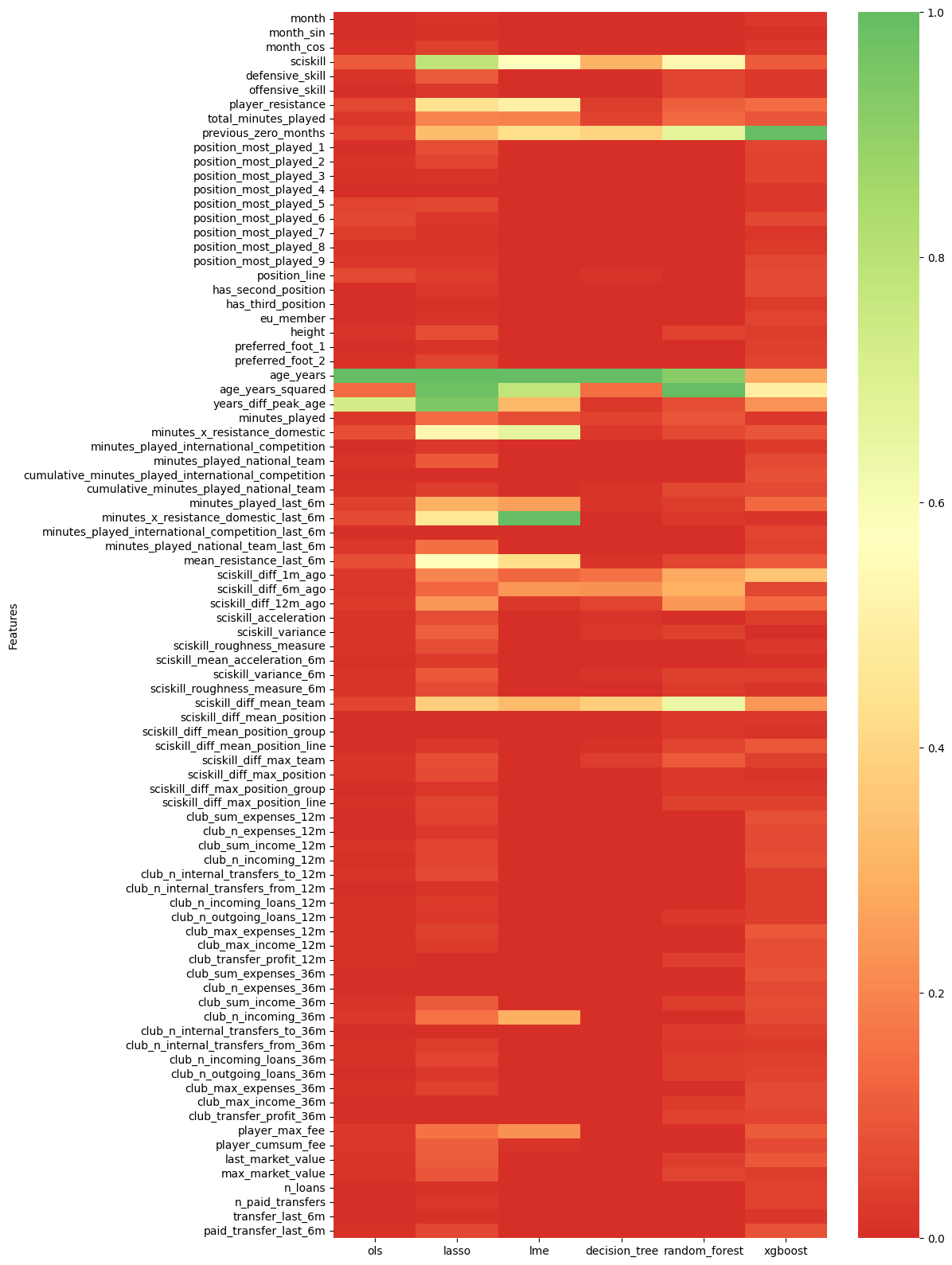}
    \end{adjustwidth}
    \caption{The feature importances of the linear and tree-based models in the prediction problem of player quality. Min--max scaling has been applied to the feature importances.}
    \label{fig:res:sciskill: feature importances}
\end{figure}

The feature importances of the linear and tree-based models are shown in Figure~\ref{fig:res:sciskill: feature importances}. The feature importances indicate that the age (`age\_years'), age squared (`age\_years\_squared'), and the difference between the age and the peak age (`years\_diff\_peak\_age') are important features amongst the models. \added[comment={Comment 11 (Reviewer B)}]{These features are strongly correlated, as they all capture age-related effects. As a consequence, the models often identify at most two of these features as important. The importance of age-related features is in line with domain knowledges since i}\deleted{I}t is known that young players tend to improve in quality, whereas older players often decrease in quality. Therefore, it is reasonable that these age-related factors are indeed important in the development of the quality of professional football players. 

\begin{figure}[H]

{\captionsetup{justification=centering}
\begin{adjustwidth}{-\extralength}{0cm}
\centering

\subfloat[\label{fig:results:sciskill:subsets:High_quality_players}]{
    \includegraphics[width=8.5cm]{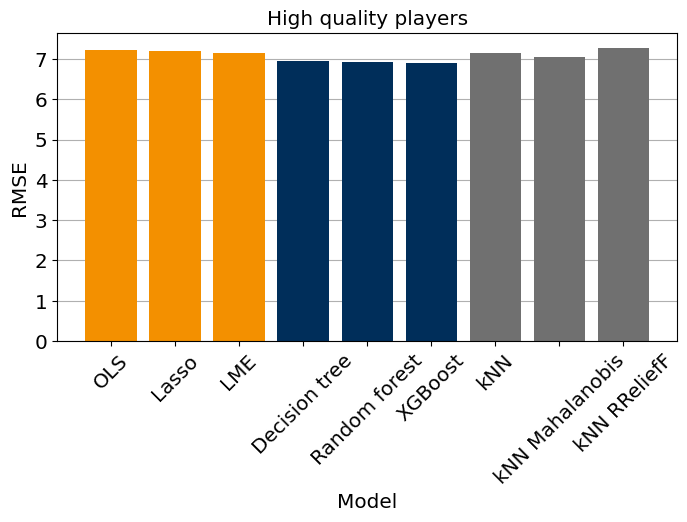}
}\\

\subfloat[\label{fig:results:sciskill:subsets:decrease}]{
    \includegraphics[width=8.5cm]{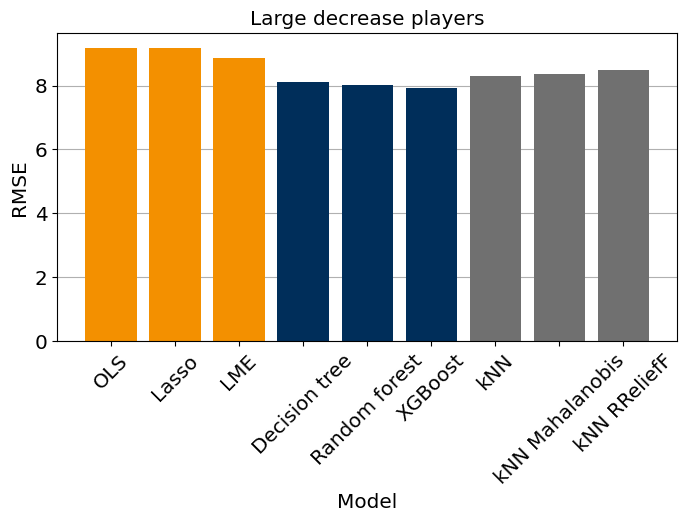}
}
\hfill
\subfloat[\label{fig:results:sciskill:subsets:increase}]{
    \includegraphics[width=8.5cm]{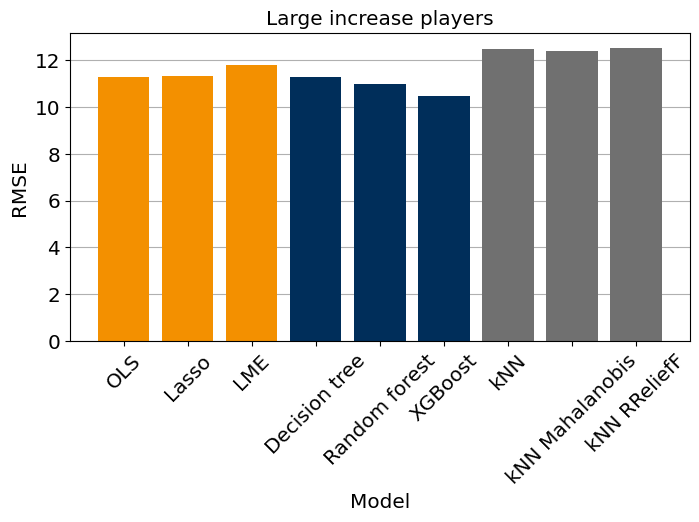}
}

\end{adjustwidth}}
\caption{
The 
 RMSE loss per model in the player quality prediction task for various player subsets: 
(\textbf{a}) players with a SciSkill of at least 100, 
(\textbf{b}) players with a SciSkill decrease of at least 10, and 
(\textbf{c}) players with a SciSkill increase of at least 10.
}
\label{fig:results:sciskill:subsets}
\end{figure}

Next to that, other features have also been assigned a large importance by the models. For instance, the player quality at prediction time (`sciskill') has a large feature importance. This indicates that the development patterns depend on the player level, which is reflected by the larger loss values in Figure~\ref{fig:results:sciskill:subsets}a compared to Figure~\ref{fig:results:sciskill:general results}a. The importance of the difference between player quality and average team quality (`sciskill\_diff\_mean\_team') can be explained by the fact that players tend to have good performances when their team plays well. Consequently, the player's quality grows towards the average team quality, which explains the importance of the difference in quality between the player and his team. Lastly, the feature importance assigns a large importance to the number of months since the last registered game (`previous\_zero\_months'). The SciSkill model penalizes players when they have not played games for a long time. As this penalty is applied after the next game of a player, the quality of a player can be expected to decrease when the number of months since the last game is large. The importance of the number of months since the last registered game shows that the models can find this pattern.


To summarize, the models indicate that the player age, player quality, the difference in quality between player and team, and the number of months since the last game are the most important features to predict future player quality.
In this way, the feature importance can be used to identify what factors are important for the development of professional football players and to test whether models behave as expected.

\subsubsection{Predictive Accuracy}
\added[comment={General Comment 1 (Reviewer 2) \& Comment on lines 600-607 (Reviewer 2)}]{The results show that the tree-based models tend to attain the lowest loss values when predicting the development of a player's quality in this specific prediction problem. The XGBoost model seems to provide the most accurate predictions for the general population of players in the data set. Moreover, the XGBoost model attains low loss values on young players, high-quality players, and players with either a large decrease or increase in football quality in this data set. The random forest model attains similar performances on the young players and most of the interesting subgroups of players. Overall, the XGBoost model seems to have the lowest loss values on the data set for predicting player quality, followed by the random forest model.}

\deleted[]{The results of training the models to predict the future player quality show that the XGBoost model provides the most accurate predictions on the general population of players. The XGBoost model outperforms the other models on young players, high-quality players, and players with either a large decrease or increase in football quality. The random forest model predicts second best, with similar performances on the young players and most of these interesting subgroups of players. For the problem of predicting a football player's development in quality, it is concluded that the best predictive performance is attained by the XGBoost model, followed closely by the random forest model.}

\subsection{Prediction Problem on Player Value}\label{res:prediction problem player value (etv)}
\subsubsection{General Population of Players}
\begin{figure}[H]
{\captionsetup{justification=centering}
\begin{adjustwidth}{-\extralength}{0cm}
\centering

\subfloat[\label{fig:results:etv:general:rmse}]{
    \includegraphics[width=8.5cm]{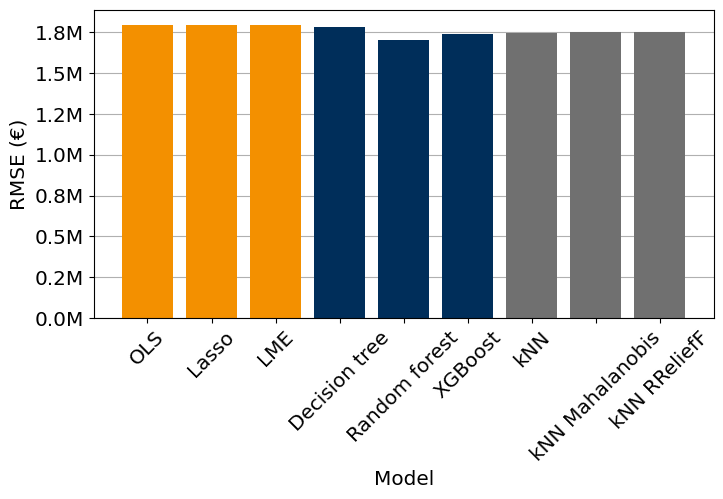}
}
\hfill
\subfloat[\label{fig:results:etv:general:mae}]{
    \includegraphics[width=8.5cm]{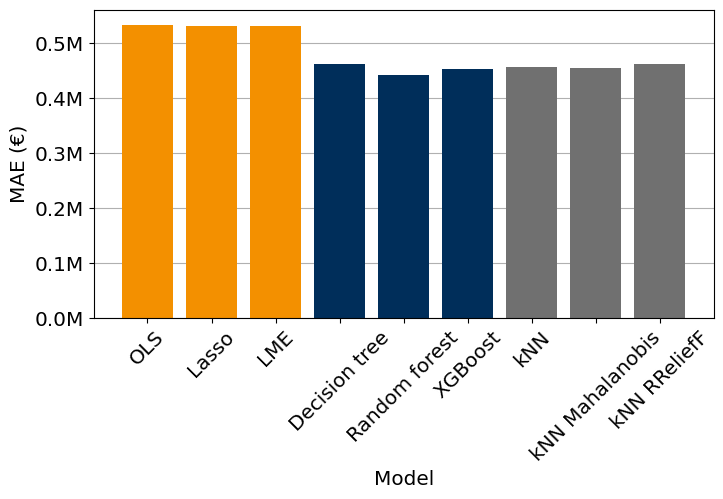}
}

\end{adjustwidth}}
\caption{
The test loss values for the different models on the general population of players in the player value prediction task for two different loss functions: 
(\textbf{a}) RMSE and
(\textbf{b}) MAE.
}
\label{fig:results:etv:general results}
\end{figure}

The loss values for predicting player value development in the prediction problem of the monetary player value are given in Figure~\ref{fig:results:etv:general results}. \added[comment={General Comment 1 (Reviewer 2)}]{The results show that the tree-based and kNN-based models perform relatively well in this prediction problem.} Although the differences in RMSE are less obvious than the differences in the MAE, the results show that the random forest model attains the lowest loss values for both the RMSE and MAE\replaced[comment={General Comment 1 (Reviewer 2)}]{, while the XGBoost and kNN-based models attain only slightly higher loss values.}{. The XGBoost model has the second-best values.}

%
%
%

The two loss functions show differences in performance as the linear models predict relatively worse with respect to the MAE than in terms of the RMSE compared to the other models. It can be reasoned that the linear models have relatively few large errors, which corresponds to the players that are harder to predict. On the contrary, the linear models predicted worse for players who are easier to predict. This means that the linear models predicted well for the players whose development is less predictable, but relatively badly for the players who are relatively easy to predict.

Additionally, the kNN models, which have time series information as features, predict better than the linear models. These results show that the time series of the player performance indicators contains important information for the development of the transfer value and that nonlinearities and interactions are involved. 

\subsubsection{Predictions per Age}
\begin{figure}[H]
    
    \includegraphics[width=0.99\linewidth]{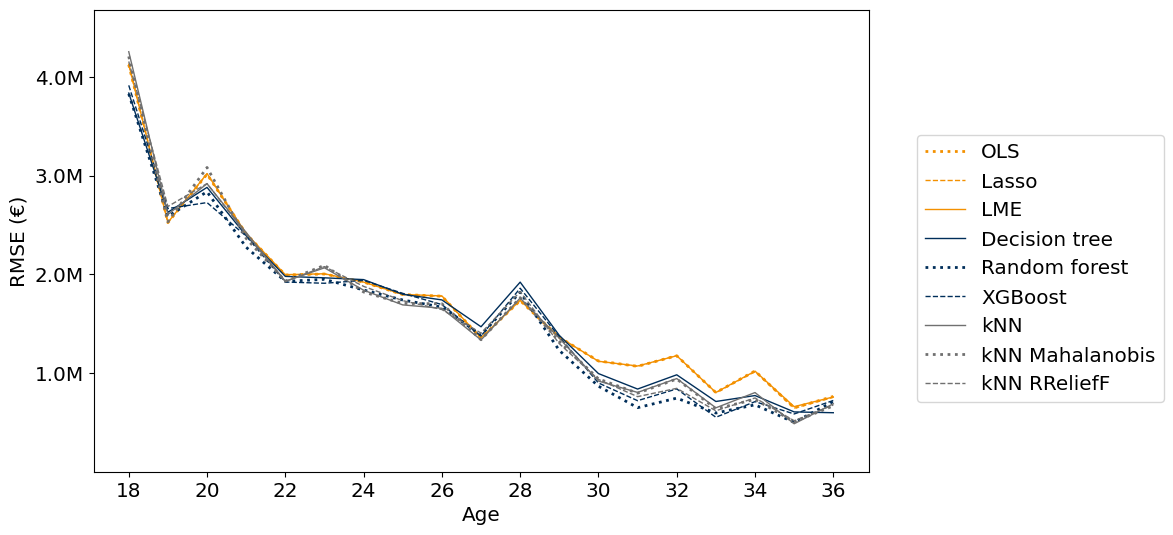}
    \caption{The RMSE values for each age for the different models in the player value prediction problem.}
    \label{fig:results:etv:error per age}
\end{figure}

The RMSE values on the test set for each age are given in Figure~\ref{fig:results:etv:error per age}. The results show that the loss values of the models are similar at all ages and \replaced[comment={General Comment 1 (Reviewer 2)}]{no clear distinction can be made between the quality of the models based on this.}{that there is no model that outperforms the other models on players of a young age. Therefore, no model is favorable based on the prediction of young players.}

Moreover, the linear models perform worse on older players. This suggests that the patterns in the development of old players involve interaction effects or nonlinearities that were not captured by the linear models.

Generally, the models tend to predict the future development of player value worse for young players and better for older players. The development of a young player is less predictable, similarly as found for the prediction of player quality in Section~\ref{res:prediction problem player quality (sciskill)}. This is due to the fact that the KPIs for young players are more volatile. On the other hand, the player values of older players are smaller and more predictable because older players tend to decrease in value. In this way, it can be explained why the models have large loss values for young players and small loss values for older players.

\subsubsection{Prediction on Important Player Groups}
\begin{figure}[H]
{\captionsetup{justification=centering}
\begin{adjustwidth}{-\extralength}{0cm}
\centering

\subfloat[\label{fig:results:etv:subsets:High_quality_players}]{
    \includegraphics[width=8.5cm]{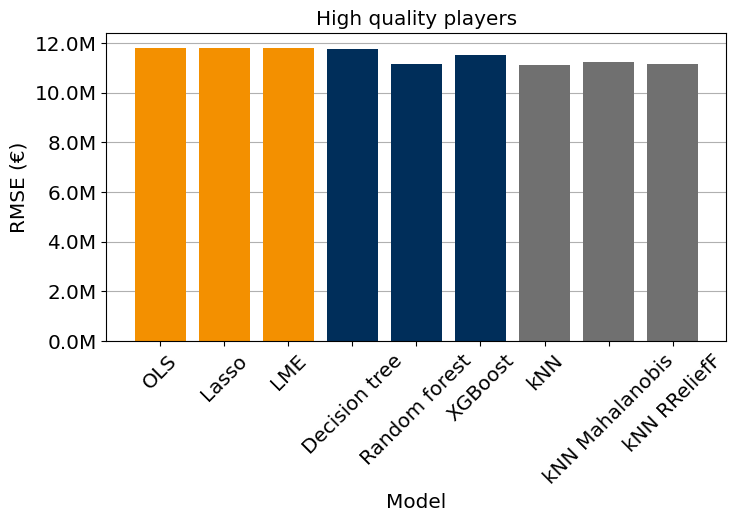}
}
\hfill
\subfloat[\label{fig:results:etv:subsets:High_value_players}]{
    \includegraphics[width=8.5cm]{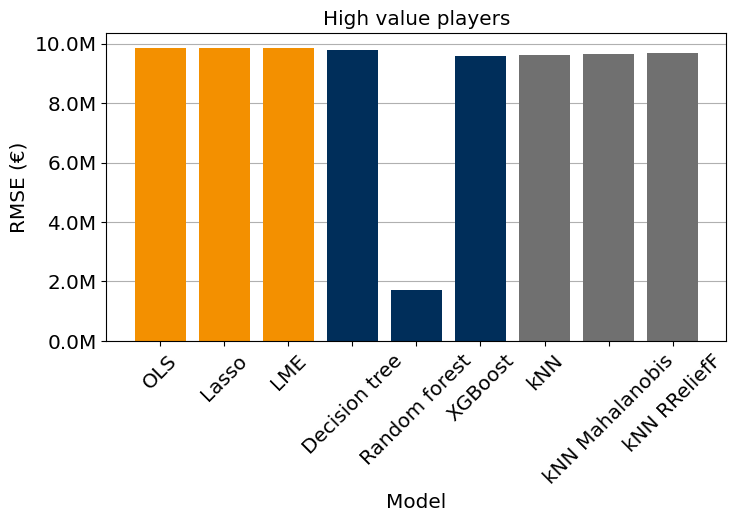}
}\\

\subfloat[\label{fig:results:etv:subsets:decrease}]{
    \includegraphics[width=8.5cm]{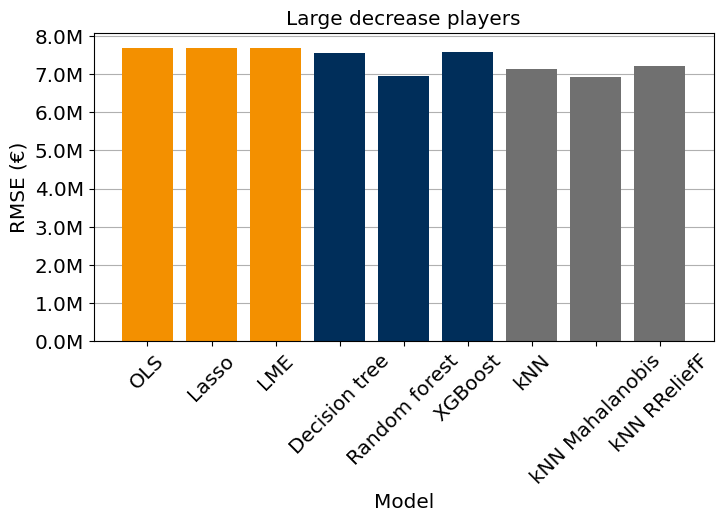}
}
\hfill
\subfloat[\label{fig:results:etv:subsets:increase}]{
    \includegraphics[width=8.5cm]{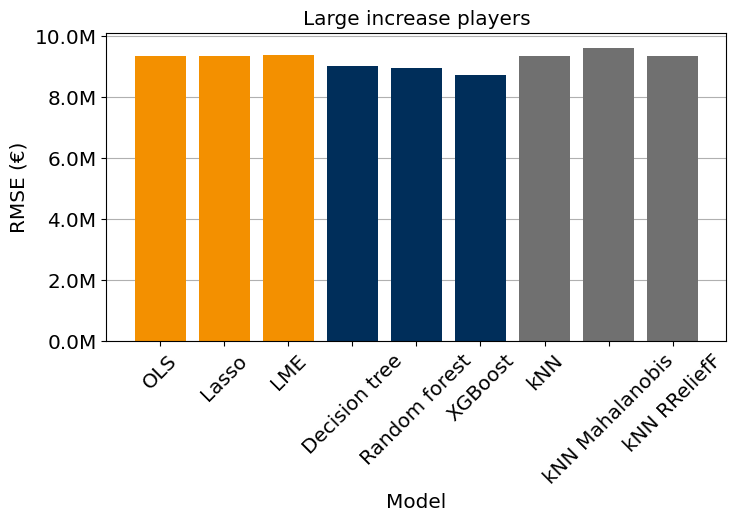}
}
\end{adjustwidth}}
\caption{
The RMSE loss per model in the player value prediction task for various player subsets: 
(\textbf{a}) players with a SciSkill of at least 100, 
(\textbf{b}) players with an ETV of at least EUR 10M, 
(\textbf{c}) players with a value decrease of at least EUR 2.5M, and 
(\textbf{d}) players with a value increase of at least EUR 2.5M.
}
\label{fig:results:etv:subsets}
\end{figure}

For the development of the transfer value, the four different \deleted{important }groups of players that have been identified as important for the application of the models are high-quality players, high-value players, players with a large decrease in transfer value, and players with a large improvement in transfer value. \added[comment={Comment 12 (Reviewer 1)}]{Based on domain knowledge, high-quality players were defined as players with a SciSkill of more than 100, while high-value players are defined as players with an ETV of at least 10 M EUR. An increase or decrease of more than EUR 2.5 M was defined as large.} The RMSE for these groups of players is visualized in Figure~\ref{fig:results:etv:subsets}. Similar to the prediction problem for player quality, the errors in the predictions are larger for these interesting groups of players. 

The results in Figure~\ref{fig:results:etv:subsets}a,c show that the random forest model and kNN-based models \replaced[comment={General Comment 1 (Reviewer 2)}]{attain the lowest loss values}{perform best} for the players of high quality and the players with a large decrease in value. The low loss values of the kNN-based models indicate that the time series features, described in Table~S6 in the Supplementary Materials, contain most of the important information to predict the development of these subsets of players. The good performance of the kNN also indicates that a local method is suitable for the prediction of these types of players. On the other hand, Figure~\ref{fig:results:etv:subsets}d shows that, for players with a large increase, the kNN models \replaced[comment={General Comment 1 (Reviewer 2)}]{attain worse loss RMSE values}{perform relatively worse} compared to the tree-based models. Similarly, the tree-based models also outperform the linear model on the subset of players with a large increase. These differences suggest that a large increase in player value is influenced by many different features and contains nonlinear patterns or interaction effects.

\pagebreak

Figure~\ref{fig:results:etv:subsets}b shows that \deleted{the performance of} the random forest model \added[comment={General Comment 1 (Reviewer 2)}]{has a distinctly lower RMSE} on the players with a high player value \deleted{is evidently better} than \deleted{that of} the other models. Players often attain high transfer values for a short period in their careers when they show peak performance and are young. Consequently, players with high transfer values can sometimes be expected to decrease in value shortly after obtaining high transfer values. The random forest model seems to be the only model that captures this pattern, as will be indicated by the high feature importance of the current transfer value in Section~\ref{subsec: etv feature importances}. This explains the \replaced[comment={General Comment 1 (Reviewer 2)}]{lower loss values}{significantly better performance} of the random forest model on the high-value players.

\deleted[comment={General Comment 1 (Reviewer 2)}]{Generally, the random forest model attains relatively low loss values on the interesting subgroups of players. This makes the random forest favorable over the other models. Additionally, the kNN-based models perform well on high-quality players and players with a large decrease, making them more suitable to predict star players on their decline.}

\subsubsection{Feature Importances}\label{subsec: etv feature importances}
\begin{figure}[H]
    
    \begin{adjustwidth}{-\extralength}{0cm}
    \centering
    \includegraphics[height=0.92\textheight]{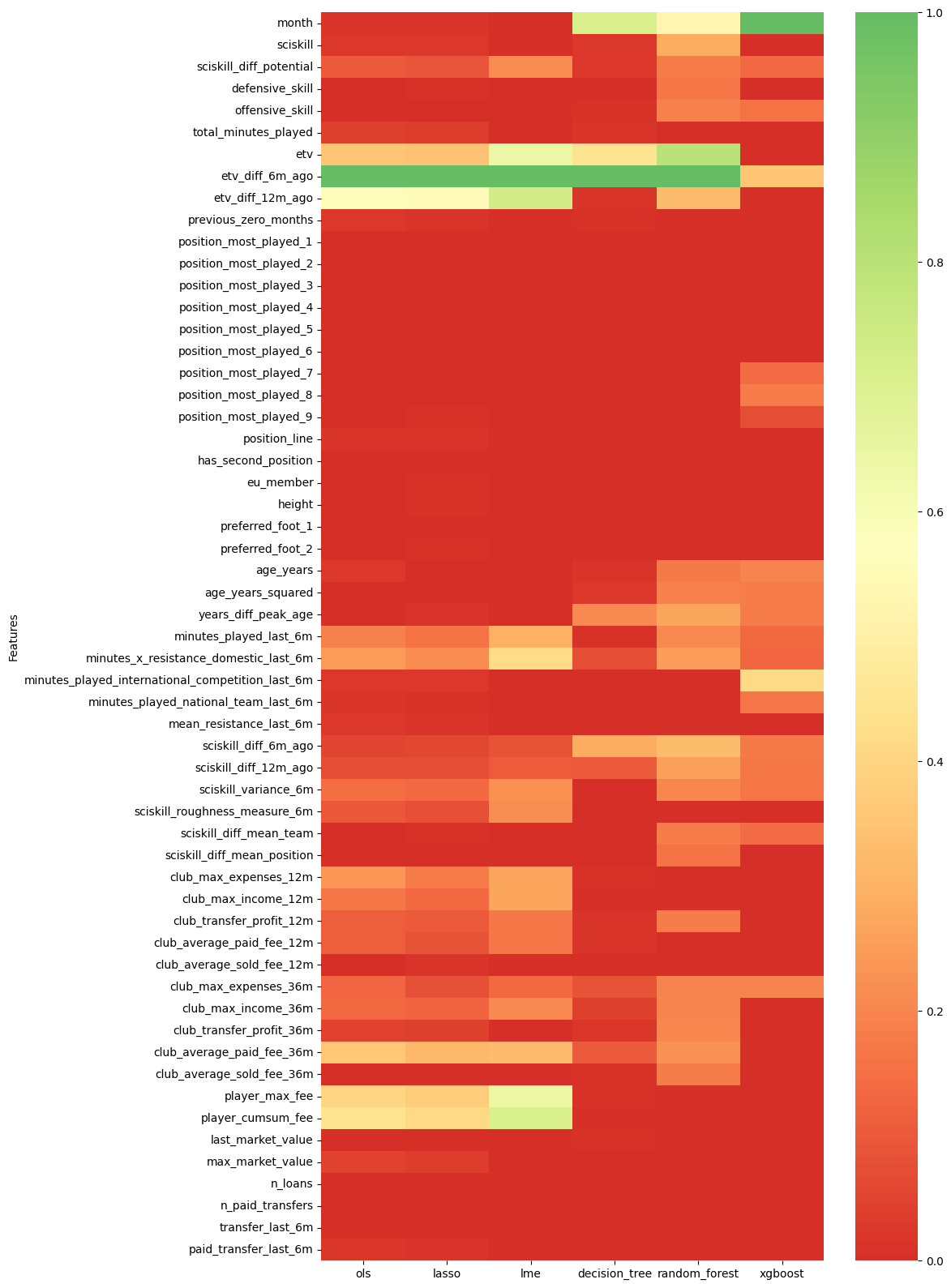}
    \end{adjustwidth}
    \caption{The 
 feature importances of the linear and tree-based models in the prediction problem of player value. Min--max scaling has been applied to the feature importances.}
    \label{fig:res:etv: feature importances}
\end{figure}

The feature importances for the development of the player transfer value after min--max scaling are shown in Figure~\ref{fig:res:etv: feature importances}. The results indicate that the most important features for predicting player value development are the features describing the most recent transfer value and the developments within the last 6 and 12 months of the player's transfer value. Especially the random forest model assigns a large value to the player value at the time of prediction (`etv'), which explains its good performance on high-value players in Figure~\ref{fig:results:etv:subsets}b. Figure~\ref{fig:res:etv: feature importances} also shows that additional time series information about the player quality (`sciskill\_diff\_6m\_ago') provides information about the development of the player transfer value. This implies that the time series information is most important in predicting the future development, which is in line with the relatively good predictive performance of the kNN models that are based on time series information.

\deleted[comment={Comment 13 (Reviewer 1)}]{Moreover, it is found that the month of the year is an important factor. This can be explained by the fact that the transfer value of a player develops differently at different times of the year. For example, a player's transfer value tends to decrease when the player only has 6 months of contract left because he could negotiate with other professional football clubs. As this commonly happens in the winter and the Estimated Transfer Value model contains the indicator whether a player has less than 6 months on his contract, the development of players' transfer value is different depending on the time of the year.}
\added[]{Moreover, it is found that the month of the year, which indicates whether it is January or July, is an important factor. Transfers in the winter transfer period are often driven by a necessity of a specific type of player at that moment, while transfers to improve the squad in the long term are more frequent in the summer transfer period. This changes the transfer fees, which is reflected in the ETV values. Additionally, players with only 6 months of contract left are able to negotiate a free transfer with a club, which decreases the transfer values of these players. As this commonly happens in the winter and the Estimated Transfer Value model contains the indicator whether a player has less than 6 months on his contract, the development of players' transfer value is different depending on the time of the year. Therefore, the fact that the month of the year is an important factor is in line with domain knowledge.}

\subsection{Predictive Accuracy}
\deleted{The results showed that the random forest model attained the best loss values on the general population of players, the groups of players with high quality, high value, or with a large increase or decrease in player value. To conclude, it is found that the random forest model provides the most accurate predictions of the development of a football player's monetary value .}
\added[comment={General Comment 1 (Reviewer 2)}]{The results showed that, with this feature set, the random forest, XGBoost, and kNN-based models attained low loss values on the general population of football players. No clear distinction was found between the RMSE values for players of different ages. The random forest and kNN models attained low RMSE scores on the subsets of high-quality players and players with a large decrease in monetary value. The tree-based models had the lowest loss values on players with a large increase in value, with the XGBoost model attaining the lowest loss. The random forest model attained a lower loss value on the subset of high-value players. Taking this into account, the random forest model has the best predictive accuracy on the data set of this prediction problem, while the XGBoost model and the kNN-based models also attain low loss values.} 

\section{Conclusions}\label{section: conclusion}
This paper aims to find the most suitable supervised learning model for forecasting the development of player performance indicators one year ahead \added[comment={General Comment 3 (Reviewer 2)}]{in a practitioner-oriented way}. Through the literature, it was found that linear regression, random forest, and kNN-based models are favorable based on their uncertainty quantification methods. 

Two prediction problems were considered to study the predictive performance of linear, tree-based, and kNN-based models. \added[comment={General Comment 1 (Reviewer 2) \& Comment on lines 600-607 (Reviewer 2)}]{The results show that the XGBoost model attained the lowest loss values for predicting the development of the player quality with this data set, while the random forest also had relatively low loss values. For the prediction of the player value, it was found that the random forest had the lowest loss values. The XGBoost and kNN-based models also had relatively accurate predictions according to some of the studied losses. }

\added[comment={Comment 14 (Reviewer 1) \& General Comment 1 (Reviewer 2) \& Comment on lines 600-607 (Reviewer 2)}]{Because the random forest had a good predictive accuracy on these data sets and it provides methods for uncertainty quantification, it seems to be a suitable model for predicting the development of player quality and value in football. Nonetheless, it is important to note that random forests can be sensitive to data imbalance, which regularly occurs in elite football. Additionally, given the large number of features needed to predict the development of football players, random forests may risk overfitting if not properly tuned and validated. These limitations should be considered when applying the model in practice, and appropriate techniques, like resampling or feature selection, may be necessary to mitigate them.}

\deleted{The results show that the predictive performance of the random forest has the second-best predictive accuracy for predicting the development of the player quality and that it provides the best predictive accuracy for the forecasting of the player value. Moreover, an off-the-shelf method for uncertainty quantification is available.}

\replaced{Taking these considerations into account, }{ Therefore,} it is concluded that the random forest is the most suitable explainable machine learning model to predict the development of player performance indicators one year ahead \added[comment={Comment on lines 600-607 (Reviewer 2)}]{for the above prediction problems}.



\section{Implications and Future Directions}\label{section: implications and future directions}

By addressing the two prediction problems, two random forest models have been obtained that can predict the development of both the quality and the monetary value of football players. These random forest models can be used to aid in transfer decisions \added[comment={General Comment 5 (Reviewer 2) \& Comment on lines 710-276 (Reviewer 2)}]{combined with a critical view of a domain expert}. Suppose the manager of a football club has a long-term interest in a football player. If the model predicting the development of player quality indicates that a player will grow in quality, it means that a player is more interesting to a football club in the long term. If the prediction of the transfer value indicates an increase in the transfer value
, it might be better to buy the player sooner rather than later. The two models can also be combined to give extra insights. Suppose a manager currently has a veteran player who is predicted to start decreasing in quality, and he can buy a young player who is predicted to develop into a first-team player within the next year. Assume that the transfer value of the young player is predicted to only slightly increase. In this case, it might be better to buy the young player one year later and then sell the veteran player \added[comment={General Comment 5 (Reviewer 2) \& Comment on lines 710-276 (Reviewer 2)}]{after a critical evaluation of the possible influences of noise and expected transfer market development by the manager. This would be beneficial} because the young player will only have a slightly higher transfer fee, and the veteran player will have a better quality in the meantime. These examples illustrate the added value of our models to predict player development for the improvement of data-informed decision-making. 

The models from this paper can also be used to complement existing methods in the literature. \citet{PantusoHvattum2021} introduced a method to optimize transfer decisions based on indicators that describe a player's quality and transfer value and their future values. Their methodology needs to know the `future' values of the player quality and their transfer values. To solve this, they consider transfer situations from over a year prior so that the values of one year later are already known. Consequently, their model can only be applied to historical situations. By using our models to predict the future values in quality and transfer value, it is possible to obtain these predictions and optimize transfer decisions with a current-time application. This makes it possible to advise data-driven transfer decisions to optimize the squad in a real-life transfer period.

\added[comment={Comment 15 (Reviewer 1) \& General Comment 2 (Reviewer 2)}]{When applying the models from this research at football clubs, possible shortcomings should be taken into consideration. The models will most likely capture patterns of development for players that are typical for the population on which they are trained. They will perform less on nontypical players, which makes it harder to predict, e.g., late bloomers. Young or injury-prone players can be considered vulnerable players who are nontypical for the general population, and the predictions of the models on such players should thus be interpreted with extra care. This highlights the value of explainable modeling, which can be complemented with interpretation techniques like the use of Shapley values by \citet{Lundberg2017shap}.}

An\added{other} advantage of the explainable models in this research is that they provide methods to gain insight into important factors for prediction. The models for the prediction problems show that this can be achieved via feature importance, similar as carried out by \citet{Baouan2022}. Because our models predict the difference between the current performance indicators and those of one year later, our research can also show the influences of the indicator itself. Our results show that the value of the indicator and the historical values of the indicators are the most important features, which adds to the knowledge obtained by \citet{Baouan2022} that describes which features are important. Additionally, time-dependent variables like the period in the year and the months without games have been found to be important features in forecasting player development. This indicates that the time series of the indicators themselves contains important information on the development of football players, although our findings also suggest that contextual information gives improved predictive performance of the models.

In short, this paper studied explainable supervised models to predict player development via performance indicators. Two prediction problems were studied in which explainable models were trained to predict the development of both player quality and player value. It was found that the random forest model is the most suitable model for forecasting player development, because of the accurate predictions for both performance indicators, combined with the method for uncertainty quantification arising from the bagging procedure.

\vspace{6pt} 

\supplementary{The following supporting information can be downloaded at: \linksupplementary{s1}, Table~S1: Dependent variable of the player quality prediction problem. Table~S2: Feature descriptions of the player quality prediction problem. Table~S3: Feature descriptions of the kNN models in the player quality prediction problem. Table~S4: Dependent variable of the player value prediction problem. Table~S5: Feature descriptions of the player value prediction problem. Table~S6: Feature descriptions of the kNN models in the player value prediction problem.}





\authorcontributions{Conceptualization, K.v.A., F.G.-S., and J.S.; methodology, K.v.A.; software, K.v.A.; validation, K.v.A.; visualization, K.v.A.; investigation, K.v.A.; data curation, K.v.A.; writing---original draft preparation, K.v.A.; writing---review and editing, K.v.A., F.G.-S., and J.S.; supervision, F.G.-S. and J.S. All authors have read and agreed to the published version of the manuscript.}

\funding{This research received no external funding.}

\institutionalreview{Not applicable.}

\informedconsent{Not applicable.}


\dataavailability{Restrictions apply to the availability of this data. Data was obtained from SciSports.}

\acknowledgments{The authors thank Geurt Jongbloed for his comments on a draft version of this paper and the company SciSports for the data, computational resources, and insight into the needs of practitioners. The authors would also like to thank the two anonymous referees who helped to improve the paper with their suggestions.}


\conflictsofinterest{The authors declare no conflicts of interest.} 
\begin{adjustwidth}{-\extralength}{0cm}

\reftitle{References}

\PublishersNote{}
\end{adjustwidth}

\begin{thebibliography}{999}

\bibitem[Yang et~al.(2024)Yang, Koenigstorfer, and
  Pawlowski]{YangKoenigstorferPawlowski2024}
Yang, Y.; Koenigstorfer, J.; Pawlowski, T.
\newblock Predicting transfer fees in professional {E}uropean football before
  and during {COVID-19} using machine learning.
\newblock {\em {E}ur. {S}port Manag. {Q}.} {\bf 2024}, {\em 24},~603--623.
\newblock {\url{https://doi.org/10.1080/16184742.2022.2153898}}.

\bibitem[McHale and Holmes(2022)]{McHaleHolmes2022}
McHale, I.G.; Holmes, B.
\newblock Estimating transfer fees of professional footballers using advanced
  performance metrics and machine learning.
\newblock {\em {E}ur. {J}. {O}per. {R}es.} {\bf 2022}, {\em 306},~389--399.
\newblock {\url{https://doi.org/10.1016/j.ejor.2022.06.033}}.

\bibitem[Rein and Memmert(2016)]{Rein2016}
Rein, R.; Memmert, D.
\newblock Big data and tactical analysis in elite soccer: Future challenges and
  opportunities for sport science.
\newblock {\em {S}pringer{P}lus} {\bf 2016}, {\em 5}, 1410.
\newblock {\url{https://doi.org/10.1186/s40064-016-3108-2}}.

\bibitem[Herold et~al.(2019)Herold, Goes, Nopp, Bauer, Thompson, and
  Meyer]{HeroldGoes2019}
Herold, M.; Goes, F.; Nopp, S.; Bauer, P.; Thompson, C.; Meyer, T.
\newblock {M}achine learning in men's professional football: {C}urrent
  applications and future directions for improving attacking play.
\newblock {\em {I}nt. {J}. {S}ports {S}cience \& {C}oaching} {\bf 2019}, {\em
  14},~798--817.
\newblock {\url{https://doi.org/10.1177/1747954119879350}}.

\bibitem[Green(2012)]{Green2012}
Green, S.
\newblock {A}ssessing the Performance of {P}remier {L}eague Goalscorers.  2012.
\newblock Available online:
  \url{https://www.statsperform.com/resource/assessing-the-performance-of-premier-league-goalscorers/}
  (accessed on 27 November 2023).

\bibitem[Eggels et~al.(2016)Eggels, van Elk, and
  Pechenizkiy]{EggelsvanElkPechenizkiy2016}
Eggels, H.; van Elk, R.; Pechenizkiy, M.
\newblock Explaining soccer match outcomes with goal scoring opportunities
  predictive analytics,  2016.
\newblock In Proceedings of the {W}orkshop on {M}achine {L}earning and {D}ata
  {M}ining for {S}ports {A}nalytics 2016, {R}iva del {G}arda, {I}taly, 19
  September 2016.

\bibitem[Anzer and Bauer(2021)]{AnzerBauer2021}
Anzer, G.; Bauer, P.
\newblock A Goal Scoring Probability Model for Shots Based on Synchronized
  Positional and Event Data in Football (Soccer).
\newblock {\em {F}ront. {S}ports {A}ct. {L}iving} {\bf 2021}, {\em 3}, 624475.
\newblock {\url{https://doi.org/10.3389/fspor.2021.624475}}.

\bibitem[Mead et~al.(2023)Mead, O'Hare, and McMenemy]{MeadOHareMcMenemy2023}
Mead, J.; O'Hare, A.; McMenemy, P.
\newblock {E}xpected goals in football: {I}mproving model performance and
  demonstrating value.
\newblock {\em {PLoS ONE}} {\bf 2023}, {\em 18}, e0282295.
\newblock {\url{https://doi.org/10.1371/journal.pone.0282295}}.

\bibitem[Hvattum and Gelade(2021)]{HvattumGelade2021}
Hvattum, L.M.; Gelade, G.A.
\newblock Comparing bottom-up and top-down ratings for individual soccer
  players.
\newblock {\em Int. J. Comput. Sci. Sport} {\bf 2021}, {\em 20},~23--42.
\newblock {\url{https://doi.org/10.2478/ijcss-2021-0002}}.

\bibitem[Rudd(2011)]{Rudd2011}
Rudd, S.
\newblock {A} Framework for Tactical Analysis and Individual Offensive
  Production Assessment in Soccer Using {M}arkov Chains,  2011.
\newblock In Proceedings of the New England Symposium on Statistics in Sports,
  Cambridge, MA, USA, 24 September 2011.

\bibitem[Van~Roy et~al.(2020)Van~Roy, Robberechts, Decroos, and
  Davis]{VanRoy2020}
Van~Roy, M.; Robberechts, P.; Decroos, T.; Davis, J.
\newblock Valuing on-the-ball actions in soccer: A critical comparison of {xT}
  and {VAEP},  2020.
\newblock In Proceedings of the 2020 {AAAI} {W}orkshop on {AI} in {T}eam
  {S}ports, Hilton Midtown, New York, NY, USA, 8 February 2020.

\bibitem[{Van Arem} and Bruinsma(2024)]{VanArem2024}
{Van Arem}, K.; Bruinsma, M.
\newblock {E}xtended {xThreat}: An explainable quality assessment method for
  actions in football using game context,  2024.
\newblock In Proceedings of the 15th {I}nternational {C}onference on the
  {E}ngineering of {S}port (ISEA 2024), Loughborough, UK, 8--11
  July 2024. {\url{https://doi.org/10.17028/RD.LBORO.27045427.V1}}.

\bibitem[Decroos et~al.(2019)Decroos, Bransen, and Davis]{Decroos2019}
Decroos, T.; Bransen, L.; Davis, J.
\newblock {Actions Speak Louder Than Goals: Valuing Player Actions in Soccer},
  2019.
\newblock In Proceedings of the 25th ACM SIGKDD International Conference on
  Knowledge Discovery \& Data Mining, Anchorage, AK, USA, 4--8 August 2019.
  {\url{https://doi.org/10.1145/3292500.3330758}}.

\bibitem[Decroos and Davis(2020)]{Decroos2020}
Decroos, T.; Davis, J.
\newblock Interpretable prediction of goals in soccer,  2020.
\newblock In Proceedings of the 2020 {AAAI} {W}orkshop on {AI} in {T}eam
  {S}ports, Hilton Midtown, New York, NY, USA, 8 February 2020.

\bibitem[Van~Haaren(2021)]{VanHaaren2021}
Van~Haaren, J.
\newblock {Why Would I Trust Your Numbers? On the Explainability of Expected
  Values in Soccer},  2021.
\newblock In Proceedings of the Workshop on {A}rtificial {I}ntelligence for
  {S}ports {A}nalytics ({AISA} 2021), {V}irtual event, 17 August 2021.
  {\url{https://doi.org/10.48550/arXiv.2105.13778}}.

\bibitem[Mendes-Neves et~al.(2022)Mendes-Neves, Meireles, and
  Mendes-Moreira]{MendesNeves2022}
Mendes-Neves, T.; Meireles, L.; Mendes-Moreira, J.
\newblock Valuing Players Over Time. \emph{arXiv} \textbf{2022}, arXiv:2209.03882.

\bibitem[S{\ae}b{\o} and Hvattum(2015)]{SeaboHvattum2015}
S{\ae}b{\o}, O.D.; Hvattum, L.M.
\newblock Evaluating the efficiency of the association football transfer market
  using regression based player ratings,  2015.
\newblock In Proceedings of the 28th {N}orsk {I}nformatikkonferanse ({NIK
  2015}), {H}\o{}gskolen i \r{A}lesund, \r{A}lesund, Norway, 23--25 November
  2015.

\bibitem[Kharrat et~al.(2020)Kharrat, McHale, and L\'{o}pez
  Pe\~na]{KharratMcHalePenaLopez2020}
Kharrat, T.; McHale, I.G.; L\'{o}pez Pe\~na, J.
\newblock Plus-minus player ratings for soccer.
\newblock {\em Eur. J. Oper. Res.} {\bf 2020}, {\em 283},~726--736.
\newblock {\url{https://doi.org/10.1016/j.ejor.2019.11.026}}.

\bibitem[Pantuso and Hvattum(2021)]{PantusoHvattum2021}
Pantuso, G.; Hvattum, L.M.
\newblock Maximizing performance with an eye on the finances: A
  chance-constrained model for football transfer market decisions.
\newblock {\em TOP} {\bf 2021}, {\em 29},~583--611.
\newblock {\url{https://doi.org/10.1007/s11750-020-00584-9}}.

\bibitem[Hvattum(2020)]{Hvattum2020}
Hvattum, L.M.
\newblock Offensive and Defensive Plus-Minus Player Ratings in Soccer.
\newblock {\em Appl. Sci.} {\bf 2020}, {\em 20}, 7345.
\newblock {\url{https://doi.org/10.3390/app10207345}}.

\bibitem[De~Bacco et~al.(2024)De~Bacco, Wang, and Blei]{DeBaccoWangBlei2024}
De~Bacco, C.; Wang, Y.; Blei, D.M.
\newblock A causality-inspired adjusted plus-minus model for player evaluation
  in team sports,  2024.
\newblock In Proceedings of the {T}hird {C}onference on {C}ausal {L}earning and
  {R}easoning ({CLeaR} 2024), {L}os {A}ngeles, {CA}, {USA}, 1--3 April 2024.

\bibitem[Wolf et~al.(2020)Wolf, Schmitt, and Schuller]{WolfSchmittSchuller2020}
Wolf, S.; Schmitt, M.; Schuller, B.
\newblock A football player rating system.
\newblock {\em J. Sports Anal.} {\bf 2020}, {\em 6},~243--257.
\newblock {\url{https://doi.org/10.3233/JSA-200411}}.

\bibitem[SciSports(2020)]{SciSkill2020}
SciSports.
\newblock {S}ci{S}kill {I}ndex---{W}hy and Hows.  2020.
\newblock Available online:
  \url{https://www.scisports.com/sciskill-index-why-and-how/\#} (accessed on 27
  November 2023).

\bibitem[Franceschi et~al.(2024)Franceschi, Brocard, Follert, and
  Gouguet]{FranceschiBrocardFollerGouguet2023}
Franceschi, M.; Brocard, J.F.; Follert, F.; Gouguet, J.J.
\newblock Determinants of football players' valuations: {A} systematic review.
\newblock {\em J. Econ. Surv.} {\bf 2024}, {\em 38},~577--600.
\newblock {\url{https://doi.org/10.1111/joes.12552}}.

\bibitem[SciSports(2024)]{ETV2020}
SciSports.
\newblock {P}layer {V}aluation {M}odel,  2024.
\newblock Available online:
  \url{https://www.scisports.com/player-valuation-model/} (accessed on 5
  December 2023).

\bibitem[Davis et~al.(2024)Davis, Bransen, Devos, Jaspers, Meert, Robberechts,
  Van~Haaren, and Van~Roy]{DaviesBransenDevos2024}
Davis, J.; Bransen, L.; Devos, L.; Jaspers, A.; Meert, W.; Robberechts, P.;
  Van~Haaren, J.; Van~Roy, M.
\newblock Methodology and evaluation in sports analytics: Challenges,
  approaches, and lessons learned.
\newblock {\em Mach. Learn.} {\bf 2024}, {\em 113},~6977--7010.
\newblock {\url{https://doi.org/10.1007/s10994-024-06585-0}}.

\bibitem[Hastie et~al.(2009)Hastie, Tibshirani, Friedman, and
  Friedman]{Hastie2009}
Hastie, T.; Tibshirani, R.; Friedman, J.H.; Friedman, J.H.
\newblock {\em The Elements of Statistical Learning: Data Mining, Inference,
  and Prediction}; Springer: Boston, MA, USA, 2009; Volume~2.

\bibitem[Tibshirani(1996)]{Tibshirani1996}
Tibshirani, R.
\newblock Regression Shrinkage and Selection Via the Lasso.
\newblock {\em J. R. Stat. Soc. Ser. B Stat. Method.} {\bf 1996}, {\em
  58},~267--288.
\newblock {\url{https://doi.org/10.1111/j.2517-6161.1996.tb02080.x}}.

\bibitem[Lindstrom and Bates(1988)]{LindstromBates1988}
Lindstrom, M.J.; Bates, D.M.
\newblock {N}ewton-{R}aphson and {EM} Algorithms for Linear Mixed-Effects
  Models for Repeated-Measures Data.
\newblock {\em J. Am. Stat. Assoc.} {\bf 1988}, {\em 83},~1014--1022.
\newblock {\url{https://doi.org/10.2307/2290128}}.

\bibitem[Chen and Guestrin(2016)]{Chen2016}
Chen, T.; Guestrin, C.
\newblock {XGBoost}: A Scalable Tree Boosting System,  2016.
\newblock In Proceedings of the 22nd ACM SIGKDD International Conference on
  Knowledge Discovery and Data Mining (KKD '16), {S}an {F}rancisco, {CA},
  {USA}, 13--17 August 2016. {\url{https://doi.org/10.1145/2939672.2939785}}.

\bibitem[Dudani(1976)]{Dudani1976}
Dudani, S.A.
\newblock The Distance-Weighted k-{N}earest-{N}eighbor Rule.
\newblock {\em IEEE Trans. Syst. Man Cybern.} {\bf 1976}, {\em
  SMC-6},~325--327.
\newblock {\url{https://doi.org/10.1109/TSMC.1976.5408784}}.

\bibitem[Robnik-Sikonja and Kononenko(1997)]{RobnikSikonjaKononenko1997}
Robnik-Sikonja, M.; Kononenko, I.
\newblock An adaptation of {R}elief for attribute estimation in regression,
  1997.
\newblock In Proceedings of the Fourteenth International Conference on Machine
  Learning, San Francisco, CA, USA, 8--12 July 1997.

\bibitem[Molnar(2019)]{Molnar2019}
Molnar, C.
\newblock {\em Interpretable Machine Learning: A Guide for Making Black Box
  Models Explainable}; Version 2019-02-21; Springer: Boston, MA, USA,  2019.
\newblock 

\bibitem[Neter et~al.(2004)Neter, Kutner, Nachtsheim, and
  Li]{NeterKutnerNachtsheimLi2004}
Neter, J.; Kutner, M.H.; Nachtsheim, C.J.; Li, W.
\newblock {\em Applied Linear Statistical Models}, 5th ed.; McGraw-Hill/Irwin:
  Boston, MA, USA, 2004.

\bibitem[Wager et~al.(2014)Wager, Hastie, and Efron]{WagerHastieEfron2014}
Wager, S.; Hastie, T.; Efron, B.
\newblock Confidence Intervals for Random Forests: The Jackknife and the
  Infinitesimal Jackknife.
\newblock {\em J. Mach. Learn. Res.} {\bf 2014}, {\em 15},~1625--1651.
\newblock {\url{https://doi.org/10.48550/arXiv.1311.4555}}.

\bibitem[Arntzen and Hvattum(2021)]{ArntzenHvattum2021}
Arntzen, H.; Hvattum, L.M.
\newblock Predicting match outcomes in association football using team ratings
  and player ratings.
\newblock {\em Stat. Model.} {\bf 2021}, {\em 21},~449--470.
\newblock {\url{https://doi.org/10.1177/1471082X20929881}}.

\bibitem[Goes et~al.(2018)Goes, Kempe, Meerhoff, and
  Lemmink]{GoesKempeMeerhoffLemmink2018}
Goes, F.; Kempe, M.; Meerhoff, L.A.; Lemmink, K.A.P.M.
\newblock Not evey pass can be an assist: A data-driven model to measure pass
  effectiveness in professional soccer matches.
\newblock {\em Big Data} {\bf 2018}, {\em 7},~57--70.
\newblock {\url{https://doi.org/10.1089/big.2018.0067}}.

\bibitem[Poli et~al.(2022)Poli, Besson, and Ravenel]{PoliBessonRavenel2022}
Poli, R.; Besson, R.; Ravenel, L.
\newblock Econometric Approach to Assessing the Transfer Fees and Values of
  Professional Football Players.
\newblock {\em Economies} {\bf 2022}, {\em 10}, 4.
\newblock {\url{https://doi.org/10.3390/economies10010004}}.

\bibitem[Herm et~al.(2014)Herm, Callsen-Bracker, and
  Kreis]{HermCallsenBrackerKreis2014}
Herm, S.; Callsen-Bracker, H.M.; Kreis, H.
\newblock When the crowd evaluates soccer players' market values: {A}ccuracy
  and evaluation attributes of an online community.
\newblock {\em Sport Manag. Rev.} {\bf 2014}, {\em 17},~484--492.
\newblock {\url{https://doi.org/10.1016/j.smr.2013.12.006}}.

\bibitem[Al-Asadi and Tasdemır(2022)]{AlAsadiTasdemir2022}
Al-Asadi, M.A.; Tasdemır, S.
\newblock Predict the Value of Football Players Using {FIFA} Video Game Data
  and Machine Learning Techniques.
\newblock {\em IEEE Access} {\bf 2022}, {\em 10},~22631--22645.
\newblock {\url{https://doi.org/10.1109/ACCESS.2022.3154767}}.

\bibitem[Steve~Arrul et~al.(2022)Steve~Arrul, Subramanian, and
  Mafas]{ArrulSubramanianMafas2022}
Steve~Arrul, V.; Subramanian, P.; Mafas, R.
\newblock Predicting the Football Players’ Market Value Using Neural Network
  Model: A Data-Driven Approach,  2022.
\newblock In Proceedings of the {2022 IEEE International Conference on
  Distributed Computing and Electrical Circuits and Electronics (ICDCECE)},
  {B}allari, {I}ndia, 23--24 April 2022,
  {\url{https://doi.org/10.1109/ICDCECE53908.2022.9792681}}.

\bibitem[Behravan and Razavi(2021)]{BehravanRazavi2021}
Behravan, I.; Razavi, S.M.
\newblock A novel machine learning method for estimating football players'
  value in the transfer market.
\newblock {\em {M}ethodol. {A}ppl.} {\bf 2021}, {\em 25},~2499--2511.
\newblock {\url{https://doi.org/10.1007/s00500-020-05319-3}}.

\bibitem[Apostolou and Tjortjis(2019)]{ApostolouTjortjis2019}
Apostolou, K.; Tjortjis, C.
\newblock Sports Analytics algorithms for performance predictions,  2019.
\newblock In Proceedings of the 10th Int. Conf. on Information, Intelligence,
  Systems and Applications (IISA 2019), {P}atras, {G}reece, 15--17 July 2019.
  {\url{https://doi.org/10.1109/IISA.2019.8900754}}.

\bibitem[Pantzalis and Tjortjis(2020)]{ChazanPantzalisTjortjis2020}
Pantzalis, V.C.; Tjortjis, C.
\newblock Sports Analytics for Football League Table and Player Performance
  Prediction,  2020.
\newblock In Proceedings of the 11th Int. Conf. on Information, Intelligence,
  Systems and Applications (IISA 2020), {P}iraeus, {G}reece, 15--17 July 2020,
  {\url{https://doi.org/10.1109/IISA50023.2020.9284352}}.

\bibitem[Giannakoulas et~al.(2023)Giannakoulas, Papageorgiou, and
  Tjortjis]{GiannakoulasPapageorgiouTjortjis2023}
Giannakoulas, N.; Papageorgiou, G.; Tjortjis, C.
\newblock Forecasting Goal Performance for Top League Football Players: A
  Comparative Study.
\newblock In \emph{Proceedings of the {A}rtificial {I}ntelligence {A}pplications and
  {I}nnovations}; Maglogiannis, I., Iliadis, L., MacIntyre, J., Dominguez, M.,
  Eds.; Springer: Cham, Switzerland, 2023; Volume 676, pp. 304--315.
\newblock {\url{https://doi.org/10.1007/978-3-031-34107-6_24}}.

\bibitem[Markopoulou et~al.(2024)Markopoulou, Papageorgiou, and
  Tjortjis]{MarkopoulouPapageorgiouTjortjis2024}
Markopoulou, C.; Papageorgiou, G.; Tjortjis, C.
\newblock Diverse Machine Learning for Forecasting Goal-Scoring Likelihood in
  Elite Football Leagues.
\newblock {\em {M}ach. {L}earn. {K}nowl. {E}xtr.} {\bf 2024}, {\em
  6},~1762--1781.
\newblock {\url{https://doi.org/10.3390/make6030086}}.

\bibitem[Barron et~al.(2018)Barron, Ball, Robins, and
  Sunderland]{BarronBallRobinsSunderland2018}
Barron, D.; Ball, G.; Robins, M.; Sunderland, C.
\newblock Artificial neural networks and player recruitment in professional
  soccer.
\newblock {\em {PLoS ONE}} {\bf 2018}, {\em 13},~e0205818.
\newblock {\url{https://doi.org/10.1371/journal.pone.0205818}}.

\bibitem[Baouan et~al.(2022)Baouan, Bismuth, Bohbot, Coustou, Lacome, and
  Rosenbaum]{Baouan2022}
Baouan, A.; Bismuth, E.; Bohbot, A.; Coustou, S.; Lacome, M.; Rosenbaum, M.
\newblock What should clubs monitor to predict future value of football
  players. \emph{arXiv} \textbf{2022}, arXiv:2212.11041.

\bibitem[Seabold and Perktold(2010)]{Seabold2010}
Seabold, S.; Perktold, J.
\newblock Statsmodels: Econometric and statistical modeling with python,  2010.
\newblock In Proceedings of the 9th {P}ython in {S}cience
  {C}onference, {A}ustin, {T}X, {USA}, 28 June--3 July 2010.

\bibitem[Pedregosa et~al.(2011)Pedregosa, Varoquaux, Gramfort, Michel, Thirion,
  Grisel, Blondel, Prettenhofer, Weiss, Dubourg, Vanderplas, Passos,
  Cournapeau, Brucher, Perrot, and {{\'E}}douard Duchesnay]{Pedregosa2011}
Pedregosa, F.; Varoquaux, G.; Gramfort, A.; Michel, V.; Thirion, B.; Grisel,
  O.; Blondel, M.; Prettenhofer, P.; Weiss, R.; Dubourg, V.;  et~al.
\newblock Scikit-learn: Machine Learning in Python.
\newblock {\em J. Mach. Learn. Res.} {\bf 2011}, {\em 12},~2825--2830.

\bibitem[Douze et~al.(2024)Douze, Guzhva, Deng, Johnson, Szilvasy, Mazar\'{e},
  Lomeli, Hosseini, and J\'{e}gou.]{Douze2024}
Douze, M.; Guzhva, A.; Deng, C.; Johnson, J.; Szilvasy, G.; Mazar\'{e}, P.E.;
  Lomeli, M.; Hosseini, L.; J\'{e}gou., H.
\newblock The {F}aiss library. \emph{arXiv} \textbf{2024}, arXiv:2401.08281.

\bibitem[Slifka and Whitton(2003)]{SikonjaKononenko2003}
Slifka, M.K.; Whitton, J.L.
\newblock Theoretical and Empirical Analysis of {ReliefF} and {RReliefF}.
\newblock {\em Mach. Learn.} {\bf 2003}, {\em 53},~23--69.
\newblock {\url{https://doi.org/10.1023/A:1025667309714}}.

\bibitem[Head et~al.(2021)Head, Kumar, Nahrstaedt, Louppe, and
  Shcherbatyi]{Louppe2016}
Head, T.; Kumar, M.; Nahrstaedt, H.; Louppe, G.; Shcherbatyi, I.
\newblock scikit-optimize/scikit-optimize,  2021. 
\newblock Available online: \url{https://doi.org/10.5281/zenodo.5565057} (accessed on 11 December 2023).


\bibitem[Lundberg and Lee(2017)]{Lundberg2017shap}
Lundberg, S.; Lee, S.I.
\newblock A Unified Approach to Interpreting Model Predictions. \emph{arXiv} \textbf{2017}, arXiv:1705.07874.

\end{thebibliography}
\end{document}